\RequirePackage{lineno} 

\pdfoutput=1 

\documentclass[11pt,a4paper]{article} 

\usepackage{jheppub}

\usepackage{preprintcover}  
\PreprintCoverPaperTitle{Measurement of the low-mass Drell--Yan differential cross section
  at $\boldsymbol{\sqrt{s}}=7$~TeV using the ATLAS detector}  
\PreprintIdNumber{CERN-PH-EP-2014-020}  
\PreprintCoverAbstract{The differential cross section for the process
  $Z/\gamma^*\rightarrow \ell\ell$ ($\ell=e,\mu$) as a function of dilepton
  invariant mass is measured in $pp$ collisions at $\sqrt{s}=7$~TeV at
  the LHC using the ATLAS detector. The measurement is performed in
  the $e$ and $\mu$ channels for invariant masses between $26$~GeV and
  $66$~GeV using an integrated luminosity of $1.6$~fb$^{-1}$ collected
  in 2011 and these measurements are combined. The analysis is
  extended to invariant masses as low as $12$~GeV in the muon
  channel using $35$~pb$^{-1}$ of data collected in 2010. The cross
  sections are determined within fiducial acceptance regions and
  corrections to extrapolate the measurements to the full kinematic
  range are provided. 
 Next-to-next-to-leading-order QCD predictions provide a significantly
 better description of the results than next-to-leading-order QCD
 calculations, unless the latter are matched to a parton shower calculation. }  
\PreprintJournalName{JHEP}  

\usepackage{lineno}

\usepackage{graphicx} 

\usepackage{atlasphysics} 

\DeclareGraphicsExtensions{.pdf}
\usepackage{subfigure}
\usepackage{setspace}

\usepackage{rotating}

\title{Measurement of the low-mass Drell--Yan differential cross section
  at $\boldsymbol{\sqrt{s}}=7$~TeV using the ATLAS detector}

\author{The ATLAS Collaboration}

\abstract{ The differential cross section for the process
  $Z/\gamma^*\rightarrow \ell\ell$ ($\ell=e,\mu$) as a function of dilepton
  invariant mass is measured in $pp$ collisions at $\sqrt{s}=7$~TeV at
  the LHC using the ATLAS detector. The measurement is performed in
  the $e$ and $\mu$ channels for invariant masses between $26$~GeV and
  $66$~GeV using an integrated luminosity of $1.6$~fb$^{-1}$ collected
  in 2011 and these measurements are combined. The analysis is
  extended to invariant masses as low as $12$~GeV in the muon
  channel using $35$~pb$^{-1}$ of data collected in 2010. The cross
  sections are determined within fiducial acceptance regions and
  corrections to extrapolate the measurements to the full kinematic
  range are provided. 
 Next-to-next-to-leading-order QCD predictions provide a significantly
 better description of the results than next-to-leading-order QCD
 calculations, unless the latter are matched to a parton shower calculation. 
}

\begin{document}
\maketitle

\section{Introduction}

The Drell--Yan (DY) process of dilepton production in hadronic
interactions~\cite{DrellYan} provides important information on the
partonic structure of hadrons which is distinct from that obtained in
deep inelastic scattering (DIS) measurements (for a recent review see
ref.~\cite{reviewpaper} and the references therein). Recent
measurements from ATLAS~\cite{ATLAS-wz,ATLAS-hmdy},
CMS~\cite{CMS-wz,CMS-wz2,Chatrchyan:2013tia} and
LHCb~\cite{LHCb-wz,Aaij:2012mda} provide further information in a new
kinematic domain. Measurements reported here are made below the mass
of the $Z$ resonance and extend to a lower invariant mass than
previous ATLAS measurements. In addition the cross sections are
normalized by luminosity rather than to the $Z$ mass peak cross
section. The data are compared to theoretical calculations of the DY
process, which can now reliably be performed at
next-to-next-to-leading-order (NNLO)
precision~\cite{Hamberg:1990np,Catani:2009sm,Catani:2007vq,FEWZ1}.
Calculations at next-to-leading-order (NLO) accuracy are also
available matched to resummations at leading-logarithm (LL) or
next-to-leading logarithm (NLL) precision~\cite{mc@nlo,powhegWZ} to
accommodate soft collinear partonic emission in the initial state. A
quantitative comparison of the data to the calculations is presented
including a QCD fit to the parton distribution functions, and a
detailed discussion of theoretical uncertainties is given.

Measurements in the region of low dilepton invariant mass,
$m_{\ell\ell}< 66$~GeV, provide complementary constraints on the
parton distribution functions (PDFs) to measurements near to the mass
of the $Z$ resonance. At low $m_{\ell\ell}$, the cross section is
dominated by the electromagnetic coupling of $q\bar{q}$ pairs to the
virtual photon ($\gamma^*$), whereas at masses near the $Z$ pole the
axial and vector weak couplings of the $q\bar{q}$ pair to the $Z$
boson dominate. Therefore the observations reported here have a
different sensitivity to up-type and down-type quarks and anti-quarks
compared to measurements near the $Z$ resonance.

The new kinematic region accessible at the LHC operating at a
centre-of-mass 
energy of $\sqrt{s}=7$~TeV and the rapidity coverage of the
ATLAS detector allow low partonic momentum fractions,  $x\sim
3\times 10^{-4}$ to $\sim 1.7 \times 10^{-3}$, to be accessed at four-momentum transfer scales, $Q$, from
$Q=m_{\ell\ell}\simeq 10$~GeV to $66$~GeV. The values of $x$ and $Q$
probed are complementary to those reached at HERA~\cite{HERAPDF1.0}.

The differential cross sections, ${\rm d}\sigma/{\rm d} m_{\ell\ell}$, are
determined within two fiducial regions of acceptance in the electron
and muon decay channels. The first
measurement, termed the {\em nominal} analysis, is conducted in the
region $26<m_{\ell\ell}<66$~GeV. The minimum muon transverse momentum requirement,
$p_{\rm T}^{\mu}$, and minimum electron transverse energy requirement, $E_{\rm T}^e$, are
 $12$~GeV. This analysis uses $1.6$~fb$^{-1}$ of data
collected in 2011, taking advantage of low-threshold triggers
available in the first part of the 2011 data taking.  This provides a
statistical uncertainty on the measurement of less than $1\%$. A
second measurement performed in the muon channel only, termed the {\em extended}
analysis, is performed in a wider kinematic region spanning
$12<m_{\ell\ell}<66$~GeV. The minimum muon transverse momentum is reduced to
$6$~GeV by  taking advantage of
the lower trigger thresholds available from an integrated
luminosity of $35$~pb$^{-1}$ collected in 2010.
 Acceptance corrections are
determined which allow the measurements to be extrapolated to the full
phase space, where no kinematic cuts are applied.  The fiducial
measurements are compared to fixed-order perturbative quantum
chromodynamic (QCD) calculations at NLO and NNLO, and NLO calculations matched to LL
parton showers using PDFs from the
MSTW~\cite{MSTW2008} collaboration. In order to assess whether the measured cross sections
can be well described with modified PDFs, a QCD fit is performed
including HERA $ep$ deep inelastic scattering data~\cite{HERAPDF1.0}.


The ATLAS detector and the data and simulation samples are described in
section~\ref{section:DataMC} as are the triggers used in the analysis. The measurement selections, procedure
and uncertainties are discussed in section~\ref{section:Selection}.
The cross-section measurements are presented in
section~\ref{section:Results} and are compared to the theoretical
predictions and QCD fits. Finally, the results are summarised in section~\ref{section:Conclusion}.

\section{Data and simulation} \label{section:DataMC}

\subsection{ATLAS detector}\label{section:Detector}

The ATLAS detector~\cite{DetectorPaper:2008} is a multi-purpose
particle physics detector with forward-backward symmetric cylindrical
geometry.\footnote{ATLAS uses a right-handed coordinate system with its origin at the nominal interaction point (IP) in the centre of the detector and the $z$-axis along the beam pipe. The $x$-axis points from the IP to the centre of the LHC ring, and the $y$-axis points upward. Cylindrical coordinates $(r,\phi)$ are used in the transverse plane, $\phi$ being the azimuthal angle around the beam pipe. The pseudorapidity is defined in terms of the polar angle $\theta$ as $\eta=-\ln\tan(\theta/2)$.} The inner detector (ID)
system is immersed in a 2~T axial magnetic field and measures the
trajectories of charged particles in the pseudorapidity range
$|\eta|<2.5$.  It consists of a semiconductor pixel detector, a silicon
microstrip detector, and a transition radiation tracker, which is also
used for electron identification.

The calorimeter system covers the pseudorapidity range
$|\eta|<4.9$. The highly segmented electromagnetic calorimeter
consists of lead absorbers with liquid argon (LAr) as active material
and covers the pseudorapidity range $|\eta|<3.2$.  In the region
$|\eta|<1.8$, a pre-sampler detector using a thin layer of LAr is used
to correct for the energy lost by electrons and photons upstream of
the calorimeter. The barrel hadronic  calorimeter is a steel and
scintillator-tile detector and is situated directly outside the
envelope of the barrel electromagnetic calorimeter. It covers a
pseudorapidity range $|\eta|<1.7$. The two endcap
hadronic calorimeters have LAr as the active material and copper
absorbers and cover a pseudorapidity range of $1.5<|\eta|<3.2$.  The forward calorimeter provides coverage of
$3.1<|\eta|<4.9$ using LAr as the active material and copper and
tungsten as the absorber material.

The muon spectrometer (MS) measures the trajectory of muons in the
large superconducting air-core toroid magnets. It covers the
pseudorapidity range $|\eta|<2.7$ and is instrumented with separate
trigger and high-precision tracking chambers arranged in three layers
with increasing distance from the interaction point.  A precision measurement
of the track coordinates in the principal bending direction of the
magnetic field is provided by drift tubes in all three layers within the pseudorapidity range
$|\eta|<2.0$. At large pseudorapidities, cathode strip chambers with
higher granularity are used in the innermost plane over
$2.0<|\eta|<2.7$. The muon trigger system, which covers the
pseudorapidity range $|\eta|<2.4$, consists of resistive plate
chambers in the barrel ($|\eta|<1.05$) and thin gap chambers in the endcap regions ($1.05<|\eta|<2.4$).

A three-level trigger system is used to select events for offline
analysis.  The level-1 trigger is implemented in hardware and
uses a subset of detector information to reduce the event rate to a
design value of at most 75 kHz. This is followed by two software-based
trigger levels, level-2 and the event filter, which together
reduce the event rate to a few hundred Hz which is recorded for offline
analysis.

\subsection{Event Triggering}\label{section:Triggers}

Events are recorded by dilepton (electron
or muon) triggers using different trigger configurations to obtain the
data in 2010 and 2011.

The 2010 data were selected by a low-threshold di-muon trigger
with a transverse momentum trigger threshold of $4$ GeV. The muons
are required to have opposite charge, originate from the same event
vertex, and satisfy $m_{\ell\ell}>0.5$~GeV. The muon trigger efficiency is
determined from a large sample of $J/\psi\rightarrow\mu\mu$ events and
is measured differentially in the transverse momentum $p^{\mu}_{\rm T}$ and
pseudorapidity of the muon, $\eta^{\mu}$.
Due to
significant charge dependence at low $p_{\rm T}^{\mu}$ and high
pseudorapidity, separate trigger efficiencies are produced for
positive and negative muons.  From these results, the efficiency of
the di-muon trigger conditions are obtained differentially in
$m_{\ell\ell}$.

The 2011 muon data were collected with a di-muon trigger with a
transverse momentum threshold of $10$~GeV. The efficiency
was determined using a tag-and-probe method on a $Z\rightarrow\mu\mu$ sample recorded using a
single-muon trigger with an $18$~GeV transverse momentum threshold.

The di-electron trigger uses calorimetric information to identify
two narrow electromagnetic energy depositions. Electron
identification algorithms use further calorimetric information on the
shower shape and
fast track reconstruction to find electron candidates with a minimum
required transverse energy of $12$~GeV. The efficiency as a function
of transverse energy and pseudorapidity of the electron is determined
using a $Z\rightarrow ee$ sample recorded using a single-electron
trigger with a $20$~GeV $E_{\rm T}$ threshold, following ref.~\cite{EGamma2011}.

\subsection{Simulation}
\label{section:MonteCarlo}

Drell--Yan signal events are simulated using {\sc
  Pythia}~$6.426$~\cite{Pythia6.4} together with
leading-order
MRST LO*~\cite{Sherstnev:2007nd} parton distribution functions. Higher-order effects are approximated by the
application of NNLO K-factors computed with the {\sc Vrap}~$0.9$
program~\cite{vrap}. {\sc Pythia}~$6.426$ is also used to simulate $Z/\gamma^*
\rightarrow\tau\tau$, $W\rightarrow\mu\nu$ and $W\rightarrow e\nu$
processes, which are scaled using NNLO K-factors.  

The {\sc Mc@nlo}~$3.42$~\cite{mc@nlo} generator is used to simulate
$t\bar{t}$ production and is also scaled to NNLO accuracy using a
K-factor~\cite{Cacciari:2011hy,Baernreuther:2012ws,Czakon:2012pz,Czakon:2012zr,Czakon:2013goa,Czakon:2011xx}. Diboson ($WW,WZ,ZZ$) production is simulated using the {\sc
  Herwig}~$6.520$~\cite{herwig} generator in conjunction with
K-factors computed at NLO precision. Since the multijet background is
difficult to simulate accurately, it is estimated using data-driven
techniques supplemented with {\sc Pythia}~$6.426$ simulation of
heavy-flavour ($b\bar{b}, c\bar{c}$) jet production.


The Monte Carlo (MC) generators are interfaced to {\sc
  Tauola}~$2.4$~\cite{Davidson:2010rw} and {\sc
  Photos}~$3.0$~\cite{Golonka:2005pn} to describe $\tau$ decays and the effects
of QED final-state radiation respectively.  Multiple $pp$ collisions
within a single bunch crossing, referred to as pile-up interactions,
are accounted for by overlaying simulated minimum-bias events
produced in {\sc Pythia} tuned to ATLAS data~\cite{mc10tune,mc11tune}.

The generated particle four-momenta are passed through the ATLAS detector
simulation~\cite{atlas_simulation}, which is
based on {\sc Geant}4~\cite{geant4}.  The simulated events are
reconstructed and selected using the same software chain as for data.
The MC samples are adjusted using factors derived from data
to reflect 
mismodellings of the lepton momentum scale and resolution, trigger
efficiency, lepton reconstruction efficiency, and isolation
efficiencies~\cite{EGamma2011,ATLASCONF2011063,ATLASCONF2012125,ATLASCONF2011046}.
No corrections are applied to the MC simulation to improve the description of
the dilepton $p_{\rm T}$ spectrum; however the influence of this effect is
assessed in section~\ref{sec:syserrors}.

Theoretical predictions of the fiducial cross sections were
computed for comparison to the measured cross sections.  {\sc Fewz}
$3.1{\rm b}2$~\cite{FEWZ1,FEWZ2,FEWZ3,FEWZ4} provides a full NLO and NNLO
calculation with higher-order electroweak (HOEW) corrections
included. To avoid double-counting with the QED final-state radiation effects
simulated with {\sc Photos}, the HOEW corrections calculated by {\sc Fewz} are
chosen to exclude this effect. The HOEW calculation uses the $G_{\mu}$ electroweak parameter
scheme~\cite{Hollik:1988ii}, in which large higher-order
corrections are already absorbed in the precisely measured muon decay constant
$G_{\mu}$. This is used as input to the electroweak calculations
together with $M_W$ and $M_Z$, the $W$
and $Z$ boson masses respectively~\cite{Beringer:1900zz}.  The
renormalisation and factorisation scales are set to
$\mu_{R}=\mu_{F}=m_{\ell\ell}$. The HOEW corrections are verified by
comparisons with calculations performed with {\sc Sanc}~\cite{SANC1,SANC2}.

The fiducial cross section is also compared to {\sc
  Powheg}~\cite{powhegWZ,powheg1,powheg2,powheg3}, which provides an
NLO prediction with a leading-log parton shower (LLPS) matched to the
matrix element calculation. It is also performed in the $G_{\mu}$
electroweak scheme, with scales $\mu_{R}=\mu_{F}=m_{\ell\ell}$. These
theoretical predictions are supplemented with HOEW
corrections which are calculated separately in {\sc Fewz} at NLO in
QCD.




\section{Experimental procedure} 
\label{section:Selection}

Events are required to be taken during stable beam condition periods
and must pass detector and data-quality requirements. At least one
vertex from a proton-proton collision, referred to a as a primary vertex, reconstructed from at least three tracks is required in
each event. Leptons produced in the Drell--Yan process are expected to
be well isolated from any energy associated with jets. The degree of
isolation for electrons is defined as the scalar sum of transverse
energy, $\sum E_{\rm T}$, of additional particles contained in a cone of size
$\Delta R=\sqrt{(\Delta\phi)^2+(\Delta\eta)^2}$ around the electron,
divided by $E^e_{\rm T}$, the transverse energy of the electron. For muons the isolation is defined using the
scalar sum of transverse momentum, $\sum p_{\rm T}$, of additional tracks
divided by $p^{\mu}_{\rm T}$, the transverse momentum of the muon. These two measures of isolation provide a good
discriminant against backgrounds arising from multijet production, where
the dominant contribution is from semileptonic decays of heavy
quarks. The electron and muon channels utilise different methods to
estimate the multijet background due to the differences between
calorimetric and track based isolation criteria, and are discussed in
detail below. The contributions from multijet processes and from
$Z/\gamma^*\rightarrow\tau\tau$ decays are the two most significant
backgrounds to the signal. Additional backgrounds can arise from
events in which a jet fakes a lepton in association with a real
lepton, for example in $W+$jet production.  Smaller background
contributions are seen from $t\bar{t}$ and diboson ($WW/WZ/ZZ$) leptonic decays
in both the nominal and extended analyses.

Asymmetric minimum lepton $E_{\rm T}$ or $p_{\rm T}$ requirements are
used in the event selections to avoid the kinematic region of
$2p_{\rm T}\sim m_{\ell\ell}$ where perturbative QCD calculations are unstable and
can lead to unphysical predictions~\cite{Frixione:1997ks}.

Due to the different kinematic ranges and detector response to
electrons and muons, the selection is optimised separately for each
channel and is described in the following.

\subsection{Electron channel}

Electrons are reconstructed using a sliding-window algorithm which
matches clusters of energy deposited in the electromagnetic
calorimeter to tracks reconstructed in the inner detector.  The
calorimeter provides the energy measurement and the track is used to
determine the angular information of the electron trajectory.
Candidates are required to be well within the tracking region,
$\left|\eta^e\right|<2.4$ of the inner detector excluding a region,
$1.37<\left|\eta^e\right|<1.52$, where the transition between the
barrel and endcap electromagnetic calorimeters is difficult to model
with the simulation. Each candidate is required to satisfy  tight
electron identification~\cite{EGamma2011} criteria. In order to
further increase the purity, calorimetric isolation is used, with
$\sum E_{\rm T}/E_{\rm T}^e$ requirements within a cone of $\Delta R = 0.4$ applied
as a function of $\eta_e$. The maximum isolation value is adjusted to
maintain a constant estimated signal efficiency of approximately $98\%$.

Candidate DY events are required to have exactly two oppositely
charged electrons with $\et^e>12\GeV$ and at least one of the electrons satisfying
$\et^e>15\GeV$. The invariant mass of the pair is required to be
between $26<m_{ee}<66\GeV$. The selection efficiencies from the
electron reconstruction, identification, calorimeter isolation and
trigger requirements are determined from $Z\rightarrow ee$,
$W\rightarrow e\nu$ and
$J/\psi\rightarrow ee$ event samples in bins of $\eta^{e}$ and
$E_{\rm T}^{e}$ using a tag-and-probe method~\cite{EGamma2011}. The MC
simulation is then corrected to
reproduce the efficiencies in the data.

Inclusive multijet production is the largest
source of background and gives rise to fake electron signatures, as
well as real electrons from the semileptonic decays of $b$ and
$c$ heavy quarks (HQ). The background contribution from $W\rightarrow
e\nu$ also gives rise to real and fake electron signatures.
Fake electrons are produced in equal number for positive and negative
charges, therefore this background can be estimated from the number of
same-sign (SS) electron pairs in the data and is subtracted from the
opposite-sign (OS) electron pair sample. The number of events with
electrons from HQ decays are reduced by a factor of three by the
isolation requirement. 

After subtracting the SS contribution and the background from
$Z/\gamma^* \rightarrow\tau\tau$, $t\bar{t}$ and diboson processes
using MC predictions, the remaining HQ contribution is estimated from
data. In these background processes the lepton isolation distribution
is expected to be the same for $ee$ and $e\mu$ pairs. Therefore the HQ contribution
in the signal region can be estimated under the assumption that the
ratio of the number of $ee$ to $e\mu$ pairs in which both leptons fail
the isolation requirement is the same as the ratio of the number of
$ee$ to $e\mu$ pairs in which both leptons pass the isolation
requirement. In this estimation procedure the muon isolation is
defined by replacing $\sum p_{\rm T}$ with $\sum E_{\rm T}$ in order to ensure
similar behaviour between $ee$ and $e\mu$ pairs.  The signal
contamination of the non-isolated $ee$ sample is about $1\%$ and is
subtracted using the MC simulation. The $e\mu$ pairs are selected from a sample of
data triggered by a muon with at least $p_{\rm T}^\mu>6\GeV$ and an electron
with $E_{\rm T}^e>10$~GeV. Leptons identified in this sample were required
to have $E_{\rm T}^{\ell}>12$~GeV and were subject to the same $E_{\rm T}$,
$\eta$ and isolation criteria for selection as the $ee$ pairs.

The background from HQ processes decreases from $15\%$ at the lowest $m_{\ell\ell}$ to
$5\%$ at the highest $m_{\ell\ell}$. The SS estimate of fake electron pairs
from jets ranges between $5\%$ and $3\%$. As a cross check the $W\rightarrow
e\nu$ background estimated from simulation is found to be $0.2\%$ of
all selected events in data. The $Z/\gamma^* \rightarrow\tau\tau$
process contribution reaches a maximum of $7\%$ at $m_{\ell\ell}\sim50$~GeV
falling to $1\%$ at low invariant mass. The $t\bar{t}$ and
diboson leptonic decays contribute $1\%$ and $0.2\%$ of the total
number of observed electron pairs, respectively.

\begin{figure}[!htb]
  \singlespacing
  \centering
    \includegraphics[width=0.48\textwidth]{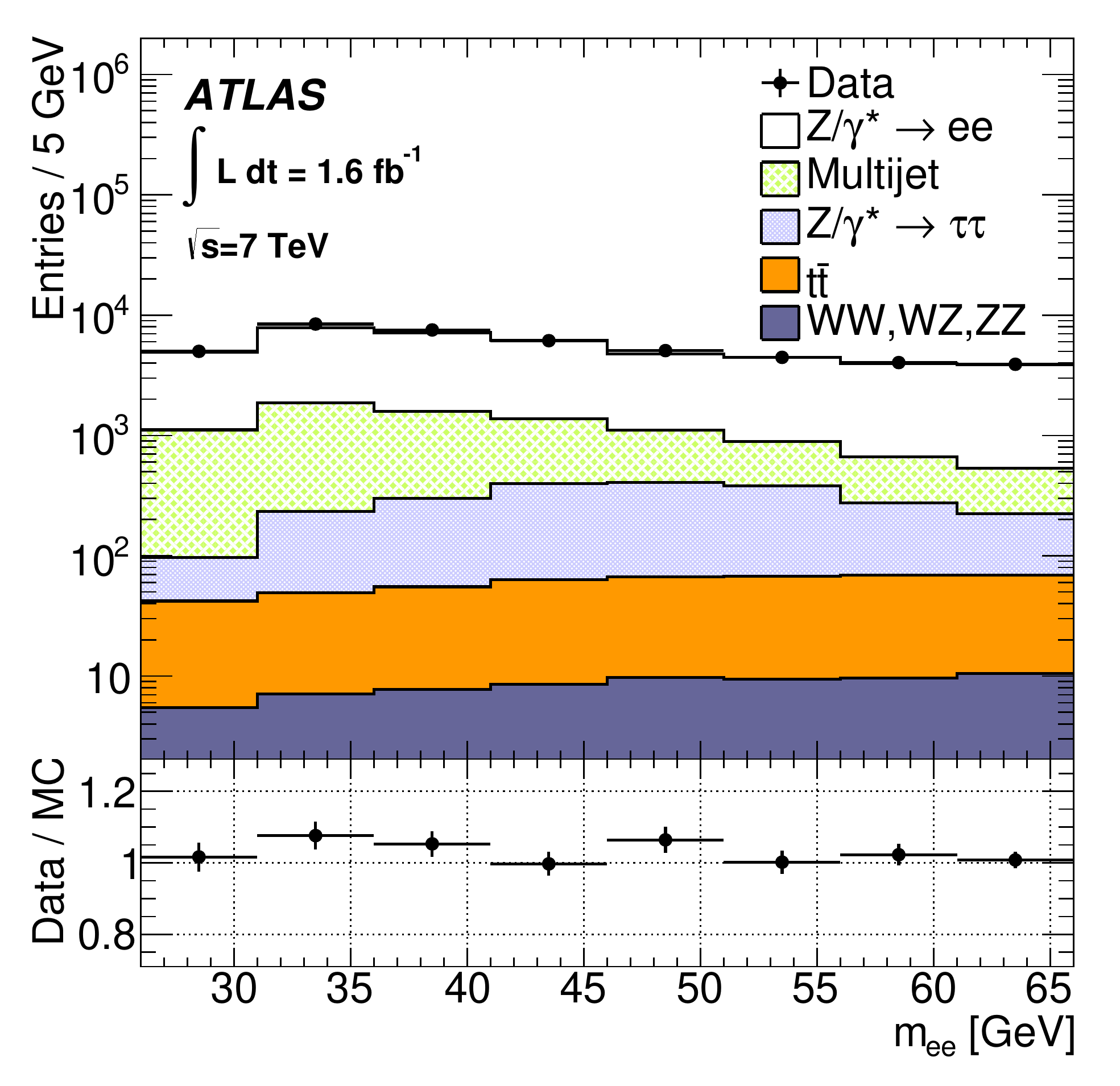}
    \caption{Distribution of di-electron invariant mass
      $m_{ee}$ for the nominal analysis selection. The error bars for the ratio represent statistical uncertainties of the data and Monte Carlo samples.
      \label{figure:ElectronControlPlots}}
\end{figure}

The invariant mass distribution, $m_{ee}$, of the final selected sample of data is
shown in figure~\ref{figure:ElectronControlPlots}, and is compared to
simulations of signal and all significant background processes.
The agreement between the data and the expected signal plus estimated
background is good.

%
%

\subsection{Muon channel}

Events in the nominal muon-channel analysis are selected online by a
trigger requiring two muon candidates with $p^{\mu}_{\rm T} > 10$ GeV.
Muons are identified
by tracks reconstructed in the muon spectrometer matched to tracks
reconstructed in the inner detector, and are required to have
$p^{\mu}_{\rm T} > 12$ GeV and $|\eta^{\mu}| < 2.4$. In addition, one
muon is required to have $p^{\mu}_{\rm T} > 15$ GeV. Isolated muons
are selected by requiring $\sum p_{\rm T}/p^{\mu}_{\rm T}<0.18$ within
a cone of $\Delta R=0.4$. The two highest $p^{\mu}_{\rm T}$ muons
should have opposite charge, and no veto on additional muons is
applied. Finally, candidate events are required to have
$26<m_{\ell\ell}<66$~GeV.
The muon track reconstruction efficiency, trigger efficiency,
isolation cut efficiency, as well as the muon momentum scale and
resolution~\cite{ATLASCONF2011046} are measured and calibrated using a
tag-and-probe method with $Z\rightarrow\mu\mu$ and
$J/\psi\rightarrow\mu\mu$ event samples. The MC
simulation is adjusted to describe the data for each of the above effects. The
trigger efficiency corrections are parameterised as a function of
$\eta^{\mu}$, $p_{\rm T}^{\mu}$ and muon charge; the muon reconstruction
efficiency corrections~\cite{ATLASCONF2011063} are determined as a function
of $\eta^{\mu}$ , $p_{\rm T}^{\mu}$ and $\phi^{\mu}$; and the muon isolation
efficiency corrections are determined as a function of $p_{\rm T}^{\mu}$ only.

The main backgrounds arise from $Z/\gamma^*\rightarrow\tau\tau$ where
the $\tau$ leptons decay leptonically to muons, and from multijet production in
which $c$- and $b$- quark mesons decay to muons.  To estimate the size
of the multijet background contribution, a partially data-driven 
two-component template fit is employed.  The multijet background is modelled
using a MC sample of heavy-flavour $b\bar{b}$ and $c\bar{c}$
jets, and a background sample obtained from SS muon pairs taken from
data.
The SS data sample accounts for any
light-flavour jets and mismodelling of the isolation spectrum. 

A pure multijet background sample is obtained by requiring all
selection cuts except the isolation cut, which is replaced by a
stringent anti-isolation requirement $\sum p_{\rm T}/p^{\mu}_{\rm T}>0.38$ placed
on one muon in the pair. The muon is chosen at random to avoid
correlations in $p^{\mu}_{\rm T}$, $\eta^{\mu}$ and charge. The second muon
then provides the shape of the multijet isolation spectrum. Templates
are constructed from the SS muon pairs in data and heavy-flavour MC simulation,
which has the MC simulation SS muon pairs subtracted to avoid double counting,
using the same anti-isolation requirement. The templates are then
fitted to the OS multijet isolation spectrum to obtain the
normalisation of each component. The procedure is validated by
comparing the isolation spectrum for data with the sum of all
background contributions after applying all selections except the
isolation cut. This method provides a significantly better
description of the isolation spectrum than MC simulation alone.

After the complete selection and application of the normalisation of
the two multijet components, the total multijet background is found to
constitute $5\%$ of the selected data sample at low $m_{\ell\ell}$
decreasing to $1\%$ at the highest $m_{\ell\ell}$, and the $Z/\gamma^*\rightarrow\tau\tau$
contribution ranges between $1\%$ and $6\%$ respectively. All remaining backgrounds together contribute less
than $1\%$ of the selected number of data events. 

\begin{figure}[!htb]
  \singlespacing
  \centering
    \includegraphics[width=0.48\textwidth]{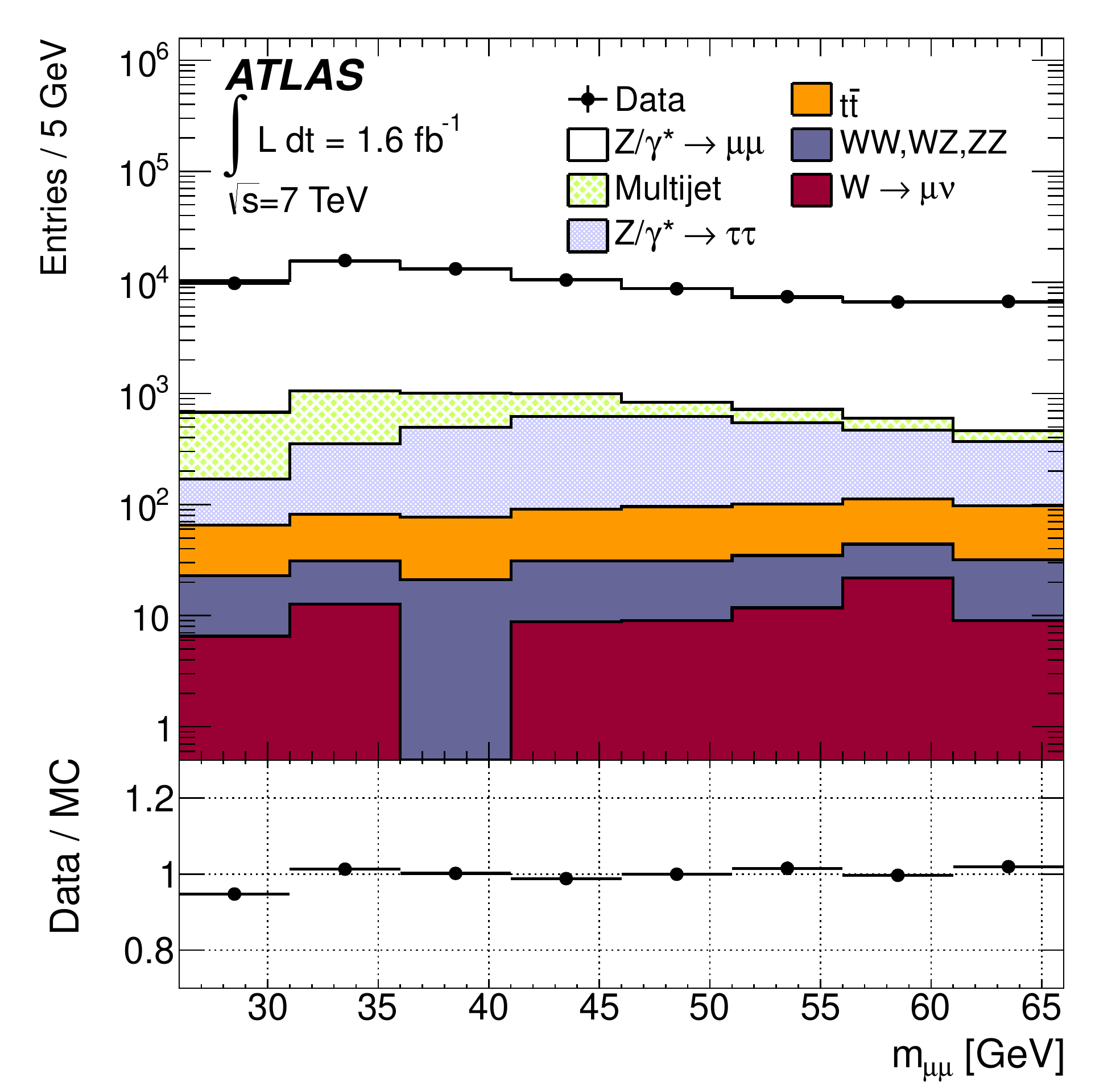}
    \caption{Distributions of  the di-muon invariant mass
      $m_{\mu\mu}$ for the nominal analysis selection. The lower panel
      shows the ratio of the data and Monte Carlo samples. The
      statistical uncertainties included in the figure are too small
      to be visible.
      \label{figure:MuonControlPlots}}
\end{figure}

Figure~\ref{figure:MuonControlPlots} shows the
distribution of invariant mass $m_{\mu\mu}$ in data after all
selections are applied. A comparison to the simulated signal
sample and all significant backgrounds is shown. Within statistical
uncertaities, a good description of the data is achieved.

%
%


\subsection{Low-mass extended analysis}

For the low-mass extended analysis, reconstructed muons are required
to have $p^{\mu}_{\rm T} > 6$ GeV and $|\eta^{\mu}| < 2.4$. In addition, one
muon is required to have $p^{\mu}_{\rm T} > 9$ GeV. Since the multijet
background is larger at lower values of $p^{\mu}_{\rm T}$, stringent
background suppression is employed by requiring the muon isolation to
be $\sum p_{\rm T}/p^{\mu}_{\rm T}<0.08$ within a cone of $\Delta R=0.6$. This
criterion is estimated from MC simulation to have a signal efficiency of $73\%$
and a background rejection of $96\%$. The reduced $p^{\mu}_{\rm T}$
requirement compared to the nominal analysis allows the measurement to
be extended to lower invariant masses. Events are required to have
$12<m_{\ell\ell}<66$~GeV.

As for the nominal muon analysis, the trigger and muon track
reconstruction efficiency, isolation cut efficiency, the muon momentum
scale and resolution are all determined using large
samples of $Z$ and $J/\psi$ decaying to $\mu\mu$. The MC simulation is
adjusted to better describe the data as detailed
elsewhere~\cite{ATLASCONF2011046,ATLASCONF2011063,MuonReco2010b}.


The largest background source comes from multijet production, which
corresponds to $23\%$ of the selected data sample at low $m_{\ell\ell}$,
falling to $6\%$ at the highest $m_{\ell\ell}$, and depends strongly
on the muon isolation requirement. The multijet contribution is
estimated using the same template fit method as for the nominal
analysis with the only difference being the anti-isolation selection
of $\sum p_{\rm T}/p^{\mu}_{\rm T}>1.0$ on one muon. All other background
contributions are at least a factor ten smaller and are estimated from
the MC samples.

\begin{figure}[!htb]
  \singlespacing
  \centering
    \includegraphics[width=0.48\textwidth]{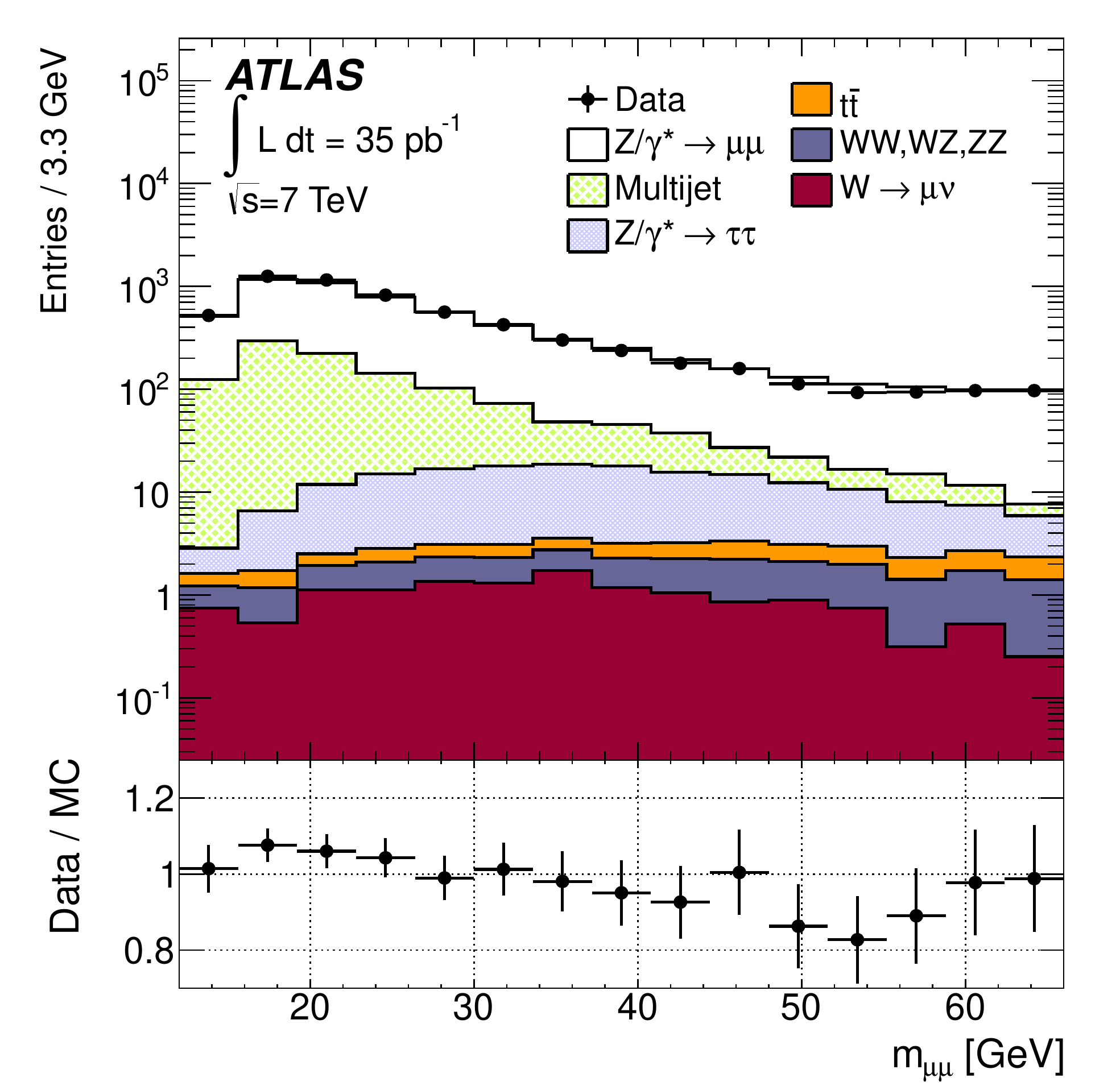}
    \caption{Distributions of the di-muon invariant mass
      $m_{\mu\mu}$ for the extended
      analysis selection. The error bars for the ratio represent statistical uncertainties of the data and Monte Carlo samples.
      \label{figure:ExtendedMuonControlPlots}}
\end{figure}

Figure~\ref{figure:ExtendedMuonControlPlots} shows the distribution
of invariant mass after all selection requirements for the data
together with
the expected contributions from all simulation samples for signal and
background processes. Within statistical uncertainties, a good
description is achieved.

\subsection{Cross-section measurement}
\label{sec:xsec}

The differential cross section ${\rm d}\sigma/{\rm d}m_{\ell\ell}$ is
determined by subtracting the estimated background from the observed
number of events and unfolding the data for detector acceptance,
selection efficiency and resolution effects using the signal
simulation samples.  The unfolding also accounts for QED final-state
radiation (FSR), and is referred to as unfolding to the Born level.  The
cross sections may also be obtained at the so-called ``dressed'' level
with respect to QED FSR, in which the leptons are recombined with any
final-state photon radiation from the leptons within a cone of $\Delta
R=0.1$ around each lepton.

The unfolded measurements are presented for a common fiducial
acceptance in the electron and muon nominal analyses within a dilepton
invariant mass of $26<m_{\ell\ell}<66$~GeV. The nominal fiducial acceptance is defined as
$|\eta^{\ell}|<2.4$, $p^{\ell}_{\rm T}>15$~GeV for the leading lepton and
$p^{\ell}_{\rm T}>12$~GeV for the sub-leading lepton, where $\ell=e\,,\mu$. The fiducial
region for the extended analysis is defined as $|\eta^{\ell}|<2.4$,
$p^{\ell}_{\rm T}>9$~GeV for the leading lepton and $p^{\ell}_{\rm T}>6$~GeV for the
sub-leading lepton, and dilepton invariant mass of
$12<m_{\ell\ell}<66$~GeV.

The muon-channel measurements are unfolded using a bin-by-bin
correction procedure. The bin widths are chosen to ensure high purity,
defined as the fraction of reconstructed signal events in a given bin
of $m_{\mu\mu}$ which were also generated in the same bin. For the
nominal analysis the bin purity is above $80\%$ in all bins, and for
the extended analysis it is always above $87\%$ due to the larger bin
widths suited to the lower integrated luminosity for this analysis.

The differential fiducial cross section for the process $pp\rightarrow
Z/\gamma^*\rightarrow \mu\mu$ is determined in each measurement bin according to

\begin{equation}
\frac{{\rm d}\sigma}{{\rm d}m_{\ell\ell}}  = \dfrac{N - B}{\mathcal{L}\,
  \mathcal{C}\,\Delta m_{\ell\ell}},  \nonumber
\label{equation:muonCrossSec}
\end{equation}
where $N$ and $B$ are the number of observed events and the estimated
number of background events respectively, $\mathcal{C}$ is
the overall signal selection efficiency and resolution smearing
correction factor determined from MC simulation, ${\mathcal{L}}$ is the
integrated luminosity of the data sample, and $\Delta m_{\ell\ell}$ is the
bin width. The $\mathcal{C}$ factor is defined as the ratio
of the number of reconstructed MC signal events passing the selection
to the number of generated MC signal events satisfying the fiducial
requirements in a given bin of $m_{\ell\ell}$ at the Born level. 
No bin centre corrections are applied for
either the muon or electron analysis.

In the electron-channel analysis the unfolding is performed using an
iterative Bayesian unfolding technique~\cite{BayesianUnfolding} due to
a bin purity of $75$\% at low $m_{\ell\ell}$, which falls to $51$\% at the
highest $m_{\ell\ell}$. The low purity at $m_{\ell\ell}\sim 60$~GeV is due to a
combination of intrinsic detector resolution, and migrations from the
$Z$ resonance peak region due to FSR  as well as the energy
loss of electrons in the material in front of the calorimeter. The
differential fiducial cross section for the process $pp\rightarrow
Z/\gamma^*\rightarrow ee$ is obtained using the relation

\begin{equation}
\frac{{\rm d}\sigma}{{\rm d}m_{\ell\ell}}  =
\frac{1}{\mathcal{L}\,\Delta m_{\ell\ell}}{\mathcal R}\dfrac{N -
  B}{\epsilon}, \nonumber
\label{equation:elecCrossSec}
\end{equation}
where $\epsilon$ includes the trigger, isolation and electron
identification efficiencies, and ${\mathcal R}$ is the response matrix.
This accounts for resolution smearing, reconstruction efficiency
effects, acceptance corrections for the region $1.37<|\eta^e|<1.52$
and unfolds the observed efficiency-corrected distribution to
the Born level.

For both the electron and muon channels, the correction from the
measured kinematics to the Born-level kinematics is
included in the $\mathcal R$ and $\mathcal{C}$ factors
respectively and can be as large as $\sim30\%$ for $m_{\ell\ell}\sim
60$~GeV due to wide-angle QED FSR radiation causing migration of events from the $Z$ resonance
peak to lower $m_{\ell\ell}$. Very good agreement in the QED FSR
predictions were found between {\sc Photos} and {\sc
  Sanc}~\cite{SANC2}.  The
corrections to the dressed level, ${\mathcal D}$, are also obtained
from MC samples, and are close to unity, but increase close to
$m_{ll}\sim 60$~GeV for the same reason since the $\Delta R=0.1$ cone
does not include all photons from wide-angle radiation.

The fiducial cross sections may be corrected to the full kinematic
range, with no lepton $p^{\ell}_{\rm T}$ or $\eta^{\ell}$ restrictions, by
applying an acceptance correction factor, ${\mathcal A}$, determined
from {\sc Fewz} and calculated at NNLO. Correction factors from the
Born to the dressed level, and from the fiducial to the full kinematic
range are provided in the cross-section tables in
section~\ref{section:Results}.
 
\subsection{Systematic uncertainties}
\label{sec:syserrors}

The systematic uncertainties on the measured cross sections are
estimated by repeating the measurement after varying each source of
uncertainty in the MC samples. The multijet background uncertainties
are determined from the comparison of different estimation methods
using data and MC simulation.

In the nominal electron channel, the detector resolution, energy
scale, and reconstruction efficiency uncertainties are propagated to
the measured cross sections by varying the MC simulation used to determine the
response matrix $\mathcal{R}$. The data and MC statistical
uncertainties are propagated by sets of pseudo-experiments where the
measured spectrum and the response matrix elements are varied by their
statistical uncertainties. The contribution to the uncertainty on the
unfolded spectrum due to the data and MC sample size, referred to as the
statistical uncertainty on the unfolding, is $1.8\%$ in the
lowest mass bin and $0.4\%$ in the highest. Variations of the electron energy
resolution and scale uncertainties yields a negligible effect. Varying the
reconstruction efficiency in the MC simulation within its uncertainty yields
correlated systematic uncertainties between mass bins ranging from
$2.3\%$ in the lowest mass bin to $0.3\%$ in the highest.
Statistical components of the systematic uncertainties are propagated
to the cross-section measurements using an ensemble of
pseudo-experiments in which replicas of the corrections are
constructed by random variation within their statistical
uncertainties.

The effects of the electron trigger, identification and isolation
requirements on the DY-pair selection efficiency are evaluated using
MC simulation by varying the correction factors that account for the
mismodellings of these selection criteria in the MC simulation.  The uncertainty on
the electron identification efficiency is partially correlated between
mass bins. The correlated component ranges from $1\%$ in the lowest
mass bin to $0.1\%$ in the highest. The uncorrelated component is
$1.2\%$ in the lowest mass bin decreasing to $0.2\%$ in the
highest. The trigger and isolation efficiency uncertainties are
treated as uncorrelated and are estimated by varying the selection
criteria used to measure these efficiencies. The uncertainty varies
from $0.1$--$0.6\%$ and $0.2$--$1.4\%$ due to the trigger and isolation
efficiencies, respectively.


The multijet background uncertainty in the nominal electron channel is
largely due to statistical uncertainties of the $e\mu$ pairs used in
the method. Systematic uncertainties are evaluated by varying the
non-isolation requirement, and the subtracted $Z/\gamma^* \rightarrow\tau\tau$ and
$t\bar{t}$ contributions in the isolated $e\mu$ sideband region by
their uncertainties. The total uncertainty on the multijet background
is approximately $15\%$, which corresponds to a fiducial cross-section
measurement uncertainty of $3.9\%$ in the lowest mass bin and $1.6\%$
in the highest. The contributions from the electroweak ($Z/\gamma^* \rightarrow\tau\tau$,
$t\bar{t}$ and diboson) backgrounds are estimated using simulation
with an uncertainty of $5\%$ on the production cross sections of
$Z/\gamma^* \rightarrow\tau\tau$ and diboson, and $6\%$ on the production cross section of
$t\bar{t}$, corresponding to an uncertainty between $0.3\%$ and
$1.0\%$ on the measured cross sections.

An uncertainty of typically less than 1\% is assigned to the cross section due
to the effect of reweighting the di-electron $p_{\rm T}$ spectrum 
to the spectrum of a different model which describes the ATLAS
measurement at the $Z$ resonance better~\cite{ATLAS-zphi}.
An uncertainty of $1.2\%$ accounts for the uncertainty in the 
{\sc Geant4} detector
simulation due to mismodelling of electron multiple scattering.

The efficiency of the muon reconstruction algorithms is well modelled
in the simulation. The uncertainty is partially correlated between
mass bins. The correlated part has a residual uncertainty of better
than $0.3\%$ over the full $\eta^{\mu}$ and $p^{\mu}_{\rm T}$ range in the
nominal analysis. For the extended analysis the uncertainty is similar
to the nominal analysis for $p^{\mu}_{\rm T}>10$~GeV, but increases to
$1.1$--$1.7\%$ at lower $p^{\mu}_{\rm T}$.

The muon momentum calibration and resolution uncertainties typically
contribute $0.5\%$ or less to the measurements in both muon-channel
analyses. The muon trigger efficiency uncertainty in the nominal
analysis is estimated by varying the $Z\rightarrow\mu\mu$ control
sample selection criteria. At low $p_{\rm T}^{\ell}$ the statistical component of
the uncertainty increases and is propagated to the cross-section
measurement using the pseudo-experiment method.  In the extended
analysis the uncertainty is dominated by the variation of the
background contribution in the $J/\psi$ resonance sample, and the
statistical sample size used to estimate the efficiency.

For the nominal muon analysis the isolation efficiency uncertainty
arises from variations of the selection criteria used to determine the
efficiency and is estimated to be $2\%$ for $p^{\mu}_{\rm T}<16$~GeV, better
than $0.5\%$ elsewhere.  In the extended analysis this uncertainty
arises from the variation of the subtracted multijet background and
the difference between two control samples used to estimate the
uncertainty.

The $Z/\gamma^* \rightarrow\tau\tau$, diboson, $W$ production, and
$t\bar{t}$ production backgrounds for both muon-channel analyses are
estimated using simulation with an uncertainty of $5\%$ on the
production cross sections, except for $t\bar{t}$ production where the
uncertainty is taken to be $6\%$.  

The multijet background uncertainty
is estimated by comparing the data and simulation in the isolation
spectrum, and by comparing $m_{\ell\ell}$ distributions for data and
simulation with an inverted isolation requirement. For the nominal
analysis the agreement in both spectra is better than $20\%$ whereas
for the extended analysis the agreement is better than
$10\%$. Variations of the background by these amounts lead to
cross-section 
uncertainties of $0.3$--$1.1\%$ for the nominal measurements,
and $0.7$--$3.0\%$ for the extended measurements where the multijet
background contribution is substantially larger. As with the electron
channel, an uncertainty is applied to the nominal muon cross section
from the reweighting of the di-muon $p_{\rm T}$ spectra, and this is seen to be
$<0.3\%$.

The uncertainty in the luminosity measurement of the ATLAS detector is fully correlated
point-to-point and also correlated between the nominal electron and
muon channel measurements. It is $1.8\%$ for the nominal analysis and
$3.5\%$ for the extended analysis~\cite{lumiPaper}. All systematic
uncertainties, including the uncertainty on the luminosity
measurement, are uncorrelated between the extended and nominal
analyses.

\section{Results}
\label{section:Results}

\subsection{Nominal analysis}

The measured Born-level fiducial cross sections for the nominal
electron analysis are presented in
table~\ref{table:electronCrossSection}, and the complete evaluation of
the individual systematic uncertainties is provided in
table~\ref{table:electronUncertainty}, excluding the normalisation
uncertainty arising from the luminosity measurement.  The sources are
separated into those which are point-to-point correlated and
uncorrelated. The precision of the electron-channel measurements is limited by the
uncertainties associated with the multijet background estimation and
the electron reconstruction efficiency. 

\begin{table}[!htb]
  \begin{center}
    \begin{tabular}{ccccc}
      \hline\hline\rule{0pt}{3ex}   
      $m_{ee}$ 	&	$\frac{\text{d}\sigma}{\text{d}m_{ee}}$	& $\delta^{\text{stat}}$&  $\delta^{\text{syst}}$ &  $\delta^{\text{total}}$    \\  \rule{0pt}{3ex} 

      [GeV] & [pb/GeV] &  [\%] & [\%] & [\%] \\
      \hline
      26--31 &	2.02	&	1.4	&	6.0	&	6.1		\\
      31--36 &	3.41	&	1.1	&	5.2	&	5.3		\\
      36--41 &	2.81	&	1.2	&	4.6	&	4.7		\\
      41--46 &	1.97	&	1.3	&	4.6	&	4.8		\\
      46--51 &	1.62	&	1.4	&	4.1	&	4.3		\\
      51--56 &	1.25	&	1.5	&	3.8	&	4.1		\\
      56--61 &	1.02	&	1.6	&	3.4	&	3.7		\\
      61--66 &	0.91	&	1.6	&	2.8	&	3.2		\\
      \hline
      \hline
    \end{tabular}
    \caption{The nominal electron-channel differential Born-level
      fiducial cross section, $\frac{\text{d}\sigma}{\text{d}m_{ee}}$. The statistical, $\delta^{\rm stat}$, systematic, $\delta^{\rm syst}$ and total, $\delta^{\rm total}$, uncertainties are given for each invariant $m_{ee}$ mass bin. The luminosity uncertainty ($1.8\%$) is not included.}
    \label{table:electronCrossSection}
  \end{center}
\end{table}

\begin{table}[!htb]
\begin{center}
\small
\begin{tabular}{c|crcccc|cccccc}
\hline\hline
& \multicolumn{6}{|c|}{Correlated} & \multicolumn{6}{c}{Uncorrelated} \\
$m_{ee}$  & $\delta^{\text{e.w.}}_{{\rm cor}}$&$\delta^{p\text{Trw}}_{{\rm cor}}$  &  $\delta^{\text{id}}_{{\rm cor1}}$ & $\delta^{\text{id}}_{{\rm cor2}}$ & $\delta^{\text{rec}}_{{\rm cor}}$ &   $\delta^{\text{geant4}}_{{\rm cor}}$ &   $\delta^{\text{trig}}$  &  $\delta^{\text{iso}}$  & $\delta^{\text{res}}_{{\rm unf}}$ & $\delta^{{\rm MC}}$ & $\delta^{\text{id}}_{{\rm unc}}$  & $\delta^{{\rm multijet}}$ \\
 
[GeV] &[\%] &[\%]&[\%]&[\%]&[\%]&[\%] &[\%]&[\%] &  [\%]& [\%] & [\%] & [\%] \\
\hline
26--31  & $-0.4$ &  0.7 & 1.0 & 0.4 & 2.3& -1.2 &  0.6 & 1.4 & 1.8 & 2.2& 1.2 & 3.9		\\
31--36  & $-0.4$ &  0.7 & 0.8 & 0.3 & 2.0& -1.2 &  0.5 & 1.1 & 1.0 &1.7& 1.1 &  3.6		 \\
36--41  & $-0.5$ &  0.6 & 0.5 & 0.2 & 1.7& -1.2 &  0.3 & 0.8 & 0.9 &1.7& 0.8 &   3.2		\\
41--46  & $-0.7$ &  1.2 & 0.3 & 0.2 & 1.4& -1.2 &  0.3 & 0.6 & 1.1 &1.9& 0.6 &   3.1		 \\
46--51  & $-0.9$ &  0.1 & 0.2 & 0.1 & 0.8& -1.2 &  0.2 & 0.4 & 1.2 &2.1& 0.4 &  2.8		\\
51--56  & $-1.0$ &  0.8 & 0.2 & 0.1 & 0.5& -1.2 &  0.2 & 0.3 & 1.4 & 2.0& 0.3 &  2.3		\\
56--61  & $-0.8$ &  0.2 & 0.1 & 0.1 & 0.4& -1.2 &  0.1 & 0.2 & 1.5 & 1.6& 0.3 &  2.0		\\
61--66  & $-0.8$ & $-0.9$ &0.1 & 0.1 & 0.3&-1.2 &  0.1 & 0.2 & 1.1 & 1.0& 0.2 &  1.6		\\ \hline
\hline
\end{tabular}
\caption{ 
The systematic uncertainties of the nominal electron-channel cross-section measurement.
Some sources of uncertainty have both correlated and uncorrelated
components. Correlated uncertainties arise from the uncertainty in the electroweak background
contributions  $\delta^{\text{e.w.}}_{{\rm cor}}$, from corrections to the Monte Carlo modelling of
the $Z/\gamma^*$ $p_{\rm T}$ spectra, $\delta^{p\text{Trw}}_{{\rm cor}}$, the electron identification efficiency,
$\delta^{\rm id}_{\rm cor1}$ and $\delta^{\rm id}_{\rm cor2}$, the reconstruction efficiency, $\delta^{\text{rec}}_{{\rm cor}}$, and from the
Geant4 simulation, $\delta^{\text{geant4}}_{{\rm cor}}$. Uncorrelated uncertainties arise from the
isolation and trigger efficiency corrections, $\delta^{\text{trig}}$ and  $\delta^{\text{iso}}$
respectively, unfolding uncertainties, $\delta^{\text{res}}_{{\rm un}}$, and the statistical
precision of the signal Monte Carlo, $\delta^{{\rm MC}}$. The electron identification
efficiency uncertainties have several components other than the two
largest correlated parts above and these are discussed in detail in
ref. \cite{EGamma2011}. These additional components are all combined into a single
uncorrelated error source $\delta^{\rm id}_{\rm unc}$. The uncertainty on the
normalisation of the multijet background is given by $\delta^{{\rm multijet}}$. The
luminosity uncertainty ($1.8\%$) is not included.
}
\label{table:electronUncertainty}
\end{center}
\end{table}


The nominal muon-channel Born-level fiducial cross section
measurements are presented in table~\ref{table:MuonCrossSection}. The
breakdown of the systematic uncertainties for the correlated and
uncorrelated sources is given in
table~\ref{table:MuonSystematics}. The precision of the measurements is limited
by the isolation efficiency determination at low $m_{\mu\mu}$.

\begin{table}[!htb]
  \begin{center}
    \begin{tabular}{ccccc}
      \hline\hline\rule{0pt}{3ex}   
      $m_{\mu\mu}$ 	&	$\frac{\text{d}\sigma}{\text{d}m_{\mu\mu}}$	& $\delta^{\text{stat}}$&  $\delta^{\text{syst}}$ &  $\delta^{\text{total}}$    \\  \rule{0pt}{3ex}	   

      [GeV] & [pb/GeV] &  [\%] & [\%] & [\%]\\
      \hline

      26--31 & 1.89 & 1.0 & 3.5 & 3.6 \\
      31--36 & 3.14 & 0.8 & 3.0 & 3.1 \\
      36--41 & 2.55 & 0.9 & 2.5 & 2.7 \\
      41--46 & 1.96 & 1.0 & 2.1 & 2.3 \\
      46--51 & 1.49 & 1.1 & 1.9 & 2.2 \\
      51--56 & 1.21 & 1.2 & 1.7 & 2.1 \\
      56--61 & 1.00 & 1.2 & 1.6 & 2.0 \\
      61--66 & 0.91 & 1.2 & 1.5 & 1.9 \\
      \hline
      \hline
    \end{tabular}
    \caption{The nominal muon-channel differential Born-level
      fiducial cross section, $\frac{\text{d}\sigma}{\text{d}m_{\mu\mu}}$. The statistical, $\delta^{\rm stat}$,
      systematic, $\delta^{\rm syst}$ and total, $\delta^{\rm total}$,
      uncertainties are given for each invariant $m_{\mu\mu}$ mass bin. The
      luminosity uncertainty ($1.8\%$) is not included.}
    \label{table:MuonCrossSection}
  \end{center}
\end{table}

\begin{table}[!htb]
  \begin{center}
  \small
    \begin{tabular}{c|ccccccc|ccccc}
      \hline \hline & \multicolumn{7}{c|}{Correlated} &
      \multicolumn{5}{c}{Uncorrelated}
      \\ $m_{\mu\mu}$&$\delta^{\text{e.w.}}$& $\delta^{p\text{Trw}}$&$\delta^{\text{reco}}_{\text{cor}}$&$\delta^{\text{trig}}_{\text{cor}}$&$\delta^{\text{iso}}_{\text{cor}}$&$\delta^{\text{multijet}}$&$\delta^{p\text{T}~\text{scale}}$
      & $\delta^{\text{trig}}_{\text{unc}}$ &
      $\delta^{\text{iso}}_{\text{unc}}$ & $\delta^{\text{res}}$&
      $\delta^{\text{MC}}$ & $\delta^{\text{reco}}_{\text{unc}}$ \\
      
      [GeV]&[\%]&[\%]&[\%]&[\%]&[\%]&[\%]&[\%]&[\%]&[\%]&[\%]&[\%]&[\%]\\ 
      
      \hline 
      26--31 & $-0.1$ & -0.2 &0.5 & 0.8 & 2.6 & $-1.1$ & $-0.5$ &  0.5 & 1.4 & 0.2 & 0.8 & 0.2 \\
      31--36 & $-0.1$ & ~0.2 &0.5 & 0.8 & 2.1 & $-1.0$ & $-0.8$ &  0.5 & 1.2 & 0.1 & 0.6 & 0.2 \\
      36--41 & $-0.2$ & ~0.0 &0.5 & 0.8 & 1.5 & $-0.8$ & $-1.0$ &  0.5 & 0.9 & 0.1 & 0.7 & 0.2 \\ 
      41--46 & $-0.3$ & -0.3 &0.5 & 0.8 & 1.1 & $-0.8$ & $-0.4$ &  0.4 & 0.7 & 0.2 & 0.8 & 0.2  \\
      46--51 & $-0.4$ & -0.1 &0.5 & 0.8 & 0.9 & $-0.5$ & $-0.6$ &  0.4 & 0.5 & 0.3 & 0.8 & 0.2  \\
      51--56 & $-0.4$ & -0.2 &0.5 & 0.7 & 0.7 & $-0.5$ & $-0.0$ &  0.3 & 0.5 & 0.2 & 0.9 & 0.2 \\ 
      56--61 & $-0.4$ & -0.2 &0.5 & 0.7 & 0.6 & $-0.4$ & $-0.3$ &  0.3 & 0.4 & 0.2 & 0.8 & 0.2  \\
      61--66 & $-0.3$ & -0.3 &0.5 & 0.6 & 0.5 & $-0.3$ & \phantom{$-$}$0.9$ &  0.2 & 0.3 & 0.2 & 0.3 & 0.2 \\
      \hline
      \hline
    \end{tabular} 
  \caption{The systematic uncertainties for the nominal muon-channel cross-section measurement. 
Some sources of uncertainty have both correlated and uncorrelated components. 
Correlated uncertainties arise from the uncertainty in the electroweak
background contributions  $\delta^{\text{e.w.}}_{{\rm cor}}$, from
corrections to the Monte Carlo modelling of the $Z/\gamma^*$ $p_{\rm  T}$ spectra, and 
from the reconstruction, trigger and isolation efficiency 
corrections, given by $\delta^{\text{reco}}_{\text{cor}}$, $\delta^{\text{trig}}_{\text{cor}}$ and $\delta^{\text{iso}}_{\text{cor}}$ respectively.
The uncertainty on the multijet and electroweak background cross
sections, given by $\delta^{\text{multijet}}$ and $\delta^{\text{e.w.}}$, respectively
and muon momentum scale uncertainty, $\delta^{p\text{T}~\text{scale}}$ are also correlated.
Uncorrelated uncertainties are due to corrections for the trigger and
isolation efficiencies,, given by $\delta^{\text{trig}}_{\text{unc}}$ and $\delta^{\text{iso}}_{\text{unc}}$
respectively. The uncertainty from the muon resolution correction, $\delta^{\text{res}}$,
from the size of the signal Monte Carlo
sample, $\delta^{\text{MC}}$, and the uncertaintities due to
corrections for the reconstruction,
$\delta^{\text{reco}}_{\text{unc}}$, are also uncorrelated. 
The luminosity uncertainty ($1.8\%$) is not included.
}
%
 \label{table:MuonSystematics}
  \end{center}
\end{table}

\begin{figure}[!htb]
  \singlespacing
  \centering
  \includegraphics[width=0.72\textwidth]{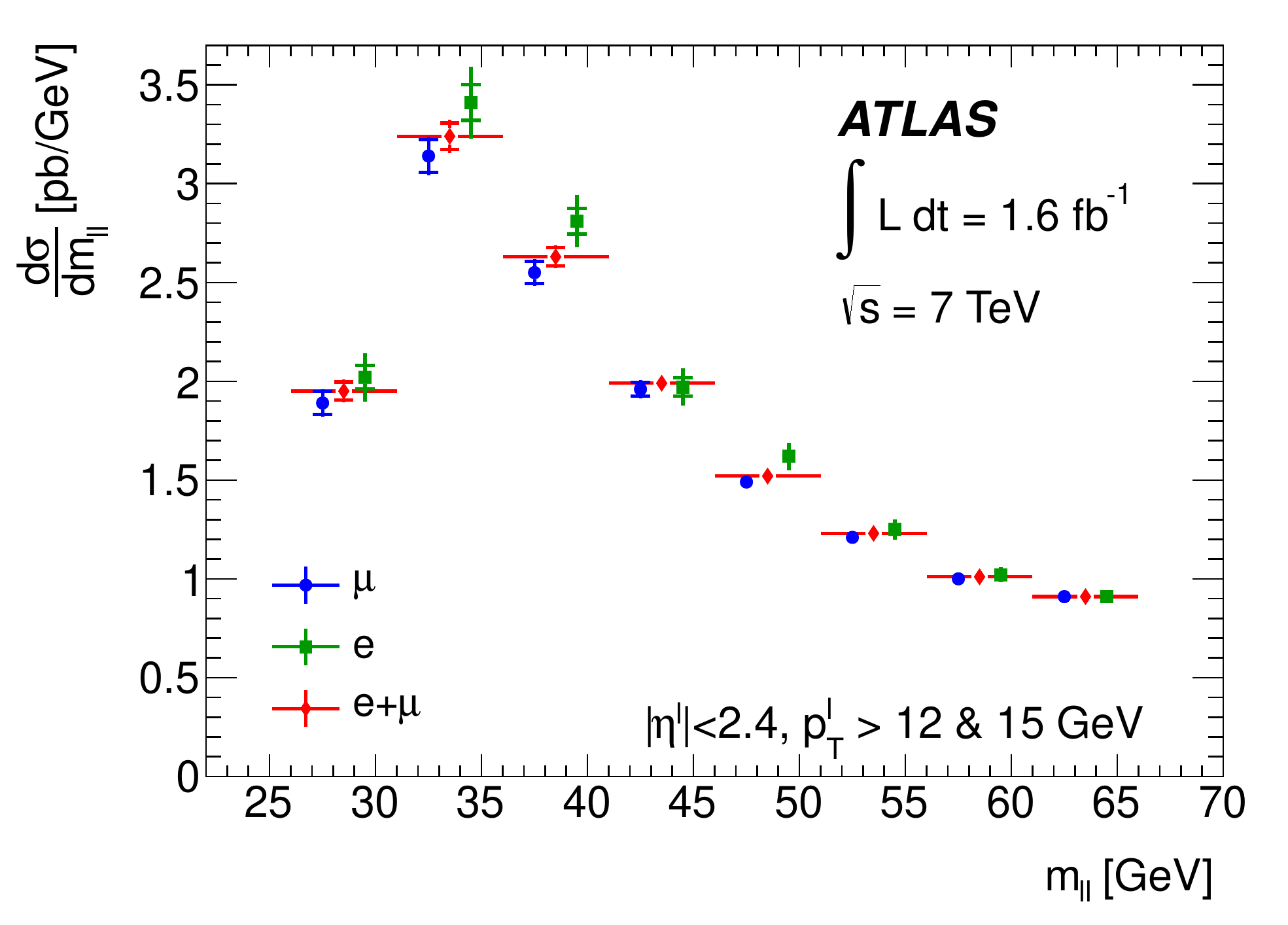}
  \caption{The fiducial Born-level combined $e$ and $\mu$ channel cross
    section as well as the individual $e$ channel and $\mu$ channel
    cross-section measurements as a function of the dilepton invariant
    mass, $m_{\ell\ell}$.
    The inner error bar represents the correlated systematic
    uncertainty and the outer error bar represents the total uncertainty in each
    bin. The electron and muon individual points are offset from
    the bin centre for the purposes of illustration. The luminosity
    uncertainty (1.8\%) is not included.}
\label{figure:CombinedFiducialCrossSection}
\end{figure}

%
Measurements made in the nominal analysis are defined with a common
fiducial acceptance and are in good agreement with each other.
A combination of the nominal $e$ and $\mu$ measurements is performed using a
$\chi^2$ minimisation technique taking into account the point-to-point
correlated systematic uncertainties of the measurements and
correlations between the electron and muon channels~\cite{Glazov:2005rn,Aaron:2009bp,HERAPDF1.0}. This method
introduces a free nuisance parameter for each correlated systematic
error source which contributes to the total $\chi^2$ and therefore
gives results that are different from a simple weighted average.  The combination
procedure yields a total $\chi^2$ per degree of freedom ($n_\text{dof}$) of
$\chi^2/n_\text{dof}=6.4/8$. There is no experimental source of systematic
uncertainty that is shifted by more than one standard deviation in the
combination. The comparison of the measured and combined cross
sections is shown in
figure~\ref{figure:CombinedFiducialCrossSection}. The electron-channel
measurements have a tendency to be larger than the muon-channel
cross sections, although they are in agreement within their
uncertainties.  The combined measurements are given in
table~\ref{tab:Comb}, which also includes the resulting correlated
uncertainty contributions after the minimisation procedure. The total
uncertainty of the cross-section measurements, excluding the
luminosity uncertainty, is reduced to $1.6$--$3\%$ across the measured
range.

\begin{sidewaystable}[!htbp]
  \begin{center}
    \small
    \resizebox{24cm}{!} {
      \begin{tabular}{c|c|cccc|ccccccccccccc|c|ccc}
        \hline \hline\rule{0pt}{3ex} 
        $m_{\ell\ell}$ &
        $\frac{\text{d}\sigma}{\text{d}m_{\ell\ell}}$ & 
        $\delta^{\text{stat}}$ &
        $\delta^{\text{cor}}$ &
        $\delta^{\text{unc}}$ & $\delta^{\text{tot}}$ & 
        $\delta^{\text{cor}}_1$ & $\delta^{\text{cor}}_2$ &
        $\delta^{\text{cor}}_3$ & $\delta^{\text{cor}}_4$ &
        $\delta^{\text{cor}}_5$ & $\delta^{\text{cor}}_6$ &
        $\delta^{\text{cor}}_7$ & $\delta^{\text{cor}}_8$ &
        $\delta^{\text{cor}}_9$ & $\delta^{\text{cor}}_{10}$ &
        $\delta^{\text{cor}}_{11}$ & $\delta^{\text{cor}}_{12}$ &
        $\delta^{\text{cor}}_{13}$ & 
        $\mathcal{D}$ &
        $\mathcal{A}$ & $\delta_{\mathcal{A}}^{\text{scale}}$ &
        $\delta_{\mathcal{A}}^{\text{pdf}+\alpha_s}$ \\
        \rule{0pt}{3ex}
        [GeV] & [pb/GeV]& [\%] & [\%] & [\%]	& [\%] & [\%]	& [\%]	& [\%] &[\%] &	[\%]& [\%] &[\%]&[\%] &	[\%]& [\%]  & [\%]   & [\%]& [\%]& & &[\%]& [\%]	\\
        \hline
        \rule{0pt}{3ex}
        
26--31& 1.95 & 0.9 & 2.4 & 1.6 &  3.0 & 0.1 & 0.4 & $-1.2$ & 0.7 & $-0.4$ & $-0.6$ & 0.4 & 0.5 & $-1.3$ & $-0.0$ & $-0.6$ & $-0.3$ & 0.8 & 0.98 & 0.069 & $^{-4.2}_{+4.2}$  &  $^{-2.0}_{+1.4}$ \\
\rule{0pt}{3ex}                     
31--36& 3.24 & 0.7 & 2.1 & 1.4 &  2.6 & 0.1 & 0.3 & $-1.1$ & 0.6 & $-0.3$ & $-0.4$ & 0.2 & 0.2 & $-1.1$ & $-0.4$ & $-0.4$ & $-0.4$ & 0.7 & 0.98 & 0.194 & $^{-2.8}_{+3.6}$  &  $^{-1.6}_{+1.1}$ \\
\rule{0pt}{3ex}                     
36--41& 2.63 & 0.8 & 1.7 & 1.2 &  2.2 & 0.2 & 0.2 & $-1.0$ & 0.5 & $-0.2$ & $-0.2$ & 0.3 & 0.3 &$-0.8$ & $-0.6$ & $-0.2$ & $-0.3$ & 0.5 & 0.99 & 0.270 & $^{-1.2}_{+1.1}$  &  $^{-1.4}_{+0.9}$ \\
\rule{0pt}{3ex}                     
41--46& 1.99 & 0.9 & 1.4 & 1.1 &  2.0 & 0.2 & 0.2 & $-1.0$ & 0.4 & $-0.2$ & $-0.0$ & 0.3 & 0.4 & $-0.5$ & $-0.2$ & $-0.2$ & $-0.0$ & 0.4 & 1.00 & 0.321 & $^{-1.2}_{+1.0}$  &  $^{-1.2}_{+0.8}$ \\
\rule{0pt}{3ex}                     
46--51& 1.52 & 0.9 & 1.2 & 1.1 &  1.9 & 0.2 & 0.3 & $-0.8$ & 0.4 & $-0.1$ &  \M$0.1$ & 0.2 & 0.3 & $-0.4$ & $-0.3$ & $-0.0$ & $-0.2$ & 0.4 & 1.05 & 0.356 & $^{-0.9}_{+0.6}$  &  $^{-1.0}_{+0.7}$ \\
\rule{0pt}{3ex}                     
51--56& 1.23 & 1.0 & 1.1 & 1.0 &  1.8 & 0.2 & 0.3 & $-0.8$ & 0.3 & $-0.1$ &  \M$0.1$ & 0.2 & 0.2 & $-0.2$ & $-0.0$ & $-0.2$ &  \M$0.1$ & 0.3 & 1.11 & 0.381 & $^{-0.4}_{+0.5}$  &  $^{-1.0}_{+0.6}$ \\
\rule{0pt}{3ex}                     
56--61& 1.01 & 1.0 & 1.0 & 1.0 &  1.7 & 0.3 & 0.3 & $-0.7$ & 0.3 & $-0.1$ &  \M$0.2$ & 0.2 & 0.2 & $-0.2$ & $-0.1$ & $-0.1$ & $-0.1$ & 0.2 & 1.19 & 0.406 & $^{-0.9}_{+0.3}$  &  $^{-0.9}_{+0.6}$ \\
\rule{0pt}{3ex}                     
61--66& 0.91 & 1.0 & 1.1 & 0.6 &  1.6 & 0.3 & 0.3 & $-0.6$ & 0.3 & $-0.0$ &  \M$0.2$ & 0.1 & 0.1 & $-0.0$ &  \M$0.7$ & $-0.1$ &  \M$0.2$ & 0.1 & 1.30 & 0.427 & $^{-0.6}_{+0.4}$  &  $^{-0.8}_{+0.5}$ \\
        \hline\hline
      \end{tabular}
    }
    \caption{The combined Born-level fiducial differential cross
      section $\frac{d\sigma}{dm_{\ell\ell}}$, statistical
      $\delta^{\text{stat}}$, total correlated $\delta^{\text{cor}}$,
      uncorrelated $\delta^{\text{unc}}$, and total
      $\delta^{\text{total}}$ uncertainties, as well as individual
      correlated sources $\delta^{\text{cor}}_i$. The correlated
      uncertainties are a linear combination of the 13 correlated
      uncertainties in the nominal muon and electron channels. As the
      uncertainties on the combined result no longer originate from
      individual error sources they are numbered 1--13. Also shown is
      the correction factor used to derive the dressed cross section
      $\left(\mathcal{D}\right)$, and the NNLO extrapolation factor
      $\left(\mathcal{A}\right)$ used to derive the cross section for
      the full phase space, along with the uncertainties associated to
      variations in scale choice
      $\delta_{\mathcal{A}}^{\text{scale}}$, and PDF uncertainty 
      $\delta_{\mathcal{A}}^{\text{pdf}+\alpha_s}$. The luminosity
      uncertainty ($1.8\%$) is not included. }
    \label{tab:Comb}
  \end{center}
\end{sidewaystable}

In addition to the combined fiducial cross sections,
table~\ref{tab:Comb} also provides two factors to obtain the 
dressed-level fiducial cross sections and to extrapolate the Born cross
sections to the full kinematic range. The former is determined by
multiplying the fiducial cross section by the dressed correction
factor $\mathcal{D}$, and the latter is determined by dividing the
fiducial cross section by the acceptance $\mathcal{A}$ as defined in
section~\ref{sec:xsec}. Both factors are obtained from MC simulation.

The acceptance correction is determined at NNLO in QCD using the {\sc Fewz}
program and is found to be sizeable at low $m_{\ell\ell}$, with a correction
factor of $0.069$ in the lowest mass bin, but increasing rapidly with
increasing $m_{\ell\ell}$. The low acceptance is largely driven by the
lepton $p_T^{\ell}$ cuts.
The calculation is subject to additional
theoretical uncertainties arising from the choice of renormalisation
and factorisation scales, $\mu_R$ and $\mu_F$ respectively, and the
choice of PDFs used in the calculation. The scales are varied
simultaneously by factors of two with respect to the default scale
choice of $\mu_R=\mu_F=m_{\ell\ell}$. The variation is taken as an
estimate of the uncertainty, which is found to be $\sim 1\%$ 
reaching $\sim4\%$ at low $m_{\ell\ell}$.
The PDF uncertainty is taken from the
MSTW2008 NNLO PDFs by taking the quadratic sum of cross-section shifts
using the $68\%$ confidence level (CL) eigenvectors and $\alpha_s$
variations~\cite{MSTW2008} and is found to be $1$--$2\%$.

\subsection{Low-mass extended analysis}

The measurements of the Born-level fiducial cross section in the
extended analysis are given in table~\ref{tab:ExtendedCrossSection},
which also includes the dressed correction factor $\mathcal{D}$, and
the acceptance $\mathcal{A}$ along with its uncertainties. The
complete breakdown of the systematic uncertainty contributions is
given in table~\ref{table:2010Uncertainties}. The dominant sources of
systematic uncertainty in this measurement are due to the trigger
efficiency and the efficiency of the isolation requirement.

\begin{table}[!htb]
  \begin{center}
    \begin{tabular}{ccccc|c|ccc}
      \hline \hline\rule{0pt}{3ex}
      $m_{\ell\ell}$ & $\frac{\text{d}\sigma}{\text{d}m_{\ell\ell}}$ & $\delta^{\text{stat}}$ &  
      $\delta^{\text{syst}}$ &  $\delta^{\text{tot}}$& $\mathcal{D}$ & $\mathcal{A}$ & 
      $\delta_{\mathcal{A}}^{\text{scale}} $ & $\delta_{\mathcal{A}}^{\text{pdf}+\alpha_s}$ \\ 

\rule{0pt}{3ex}
      [GeV] & [pb/GeV]	& [\%]	& [\%]	& [\%] & & &[\%]	& [\%]	\\ 
      \hline
      \rule{0pt}{3ex}
      12--17 &    12.41  &4.2  &12.6 & 13.3  & 1.00  &  0.04&$^{-7.1}_{+7.5}$   &$^{-4.1}_{+2.7}$     \\
      \rule{0pt}{3ex}                                 
      17--22 &    22.57  &3.1  &12.3 & 12.7  & 0.98  &  0.20&$^{-3.7}_{+4.2}$   &$^{-3.0}_{+2.0}$     \\
      \rule{0pt}{3ex}                                 
      22--28 &    14.64  &3.3  & 9.5 & 10.0  & 0.98  &  0.30&$^{-0.4}_{+0.8}$   &$^{-2.3}_{+1.6}$     \\
      \rule{0pt}{3ex}                                 
      28--36 &     6.73  &4.0  & 7.4 &  8.5  & 0.99  &  0.35&$^{-0.3}_{+0.3}$  &$^{-1.8}_{+1.2}$     \\
      \rule{0pt}{3ex}
      36--46 &     2.81  &5.2  & 5.7 &  7.8  & 1.02  &  0.39&$^{-0.3}_{+0.4}$   &$^{-1.3}_{+0.9}$     \\
      \rule{0pt}{3ex}
      46--66 &     1.27  &4.7  & 5.2 &  7.1  & 1.16  &  0.43&$^{-0.4}_{+0.7}$   &$^{-1.0}_{+0.6}$     \\
      \hline \hline
    \end{tabular}
    \caption{The extended muon channel Born-level fiducial
      differential cross section $\frac{\text{d}\sigma}{\text{d}m_{\ell\ell}}$, with
      the statistical $\delta^{\text{stat}}$, systematic
      $\delta^{\text{syst}}$, and total $\delta^{ \text{tot}}$
      uncertainties for each invariant mass bin. Also shown is the correction factor used to
      derive the dressed cross section $\left(\mathcal{D}\right)$, and
      the extrapolation factor $\left(\mathcal{A}\right)$ used to
      derive the cross section for the full phase space, along with
      the uncertainties associated to variations in scale
      $\delta_{\mathcal{A}}^{\text{scale}} $, and PDF uncertainty
      $\delta_{\mathcal{A}}^{\text{pdf}+\alpha_s}$.   The luminosity
      uncertainty ($3.5\%$) is not included. }
    \label{tab:ExtendedCrossSection}
  \end{center}
\end{table}

\begin{table}[!htb]
  \begin{center}
    \begin{tabular}{c|ccccc|cc}
      \hline \hline
     & \multicolumn{5}{c|}{Correlated} & \multicolumn{2}{c}{Uncorrelated} \\
      $m_{\mu\mu}$&  $\delta^{\text{reco}} $ & $\delta^{\text{trig}} $  &
      $\delta^{\text{iso}} $ &$\delta^{\text{multijet}} $ & $\delta^{p\text{T}\text{\,scale}}$& $\delta^{\text{res}} $ & $\delta^{\text{MC}} $\\

	[GeV]&[\%]&[\%]&[\%]&[\%]& [\%]&[\%]&[\%]\\

      \hline
      12--17  & 2.5 & 4.0 & 11.3 & $-3.0$ & $-0.2$  & 0.5 & 0.6\\
      17--22  & 1.4 & 3.7 & 11.3 & $-2.8$ & \M$0.1$  & 0.3 & 0.3\\
      22--28  & 0.9 & 3.6 & ~8.5 & $-1.8$ & \M$0.0$  & 0.1 & 0.4\\
      28--36  & 0.7 & 3.6 & ~6.2 & $-1.6$ & $-0.1$  & 0.2 & 0.4\\
      36--46  & 0.7 & 3.6 & ~4.2 & $-1.3$ & $-0.1$  & 0.1 & 0.5\\
      46--66  & 0.6 & 3.6 & ~3.6 & $-0.7$ & $-0.0$  & 0.1 & 0.5\\
    \hline
      \hline
    \end{tabular}
    \caption{The systematic uncertainties for the extended muon
      channel cross-section measurement in each invariant mass bin. Correlated
      uncertainties come from the reconstruction, trigger and
      isolation efficiency corrections, given by
      $\delta^{\text{reco}}$, $\delta^{\text{trig}}$ and
      $\delta^{\text{iso}}$ respectively. The uncertainty on the
      multijet background cross section, $\delta^{\text{multijet}}$
      and the uncertainty on the muon momentum scale,
      $\delta^{p\text{T}~\text{scale}}$, are also correlated across
      bins. Uncorrelated uncertainties are due to the uncertainty from
      the muon resolution correction, $\delta^{\text{res}}$, and the
     sample size of the signal Monte Carlo sample,
      $\delta^{\text{MC}}$.The luminosity uncertainty ($3.5\%$) is not
      included. }
    \label{table:2010Uncertainties}
  \end{center}
\end{table} 

\begin{figure}[!htb]
  \singlespacing
  \centering
  \includegraphics[width=0.72\textwidth]{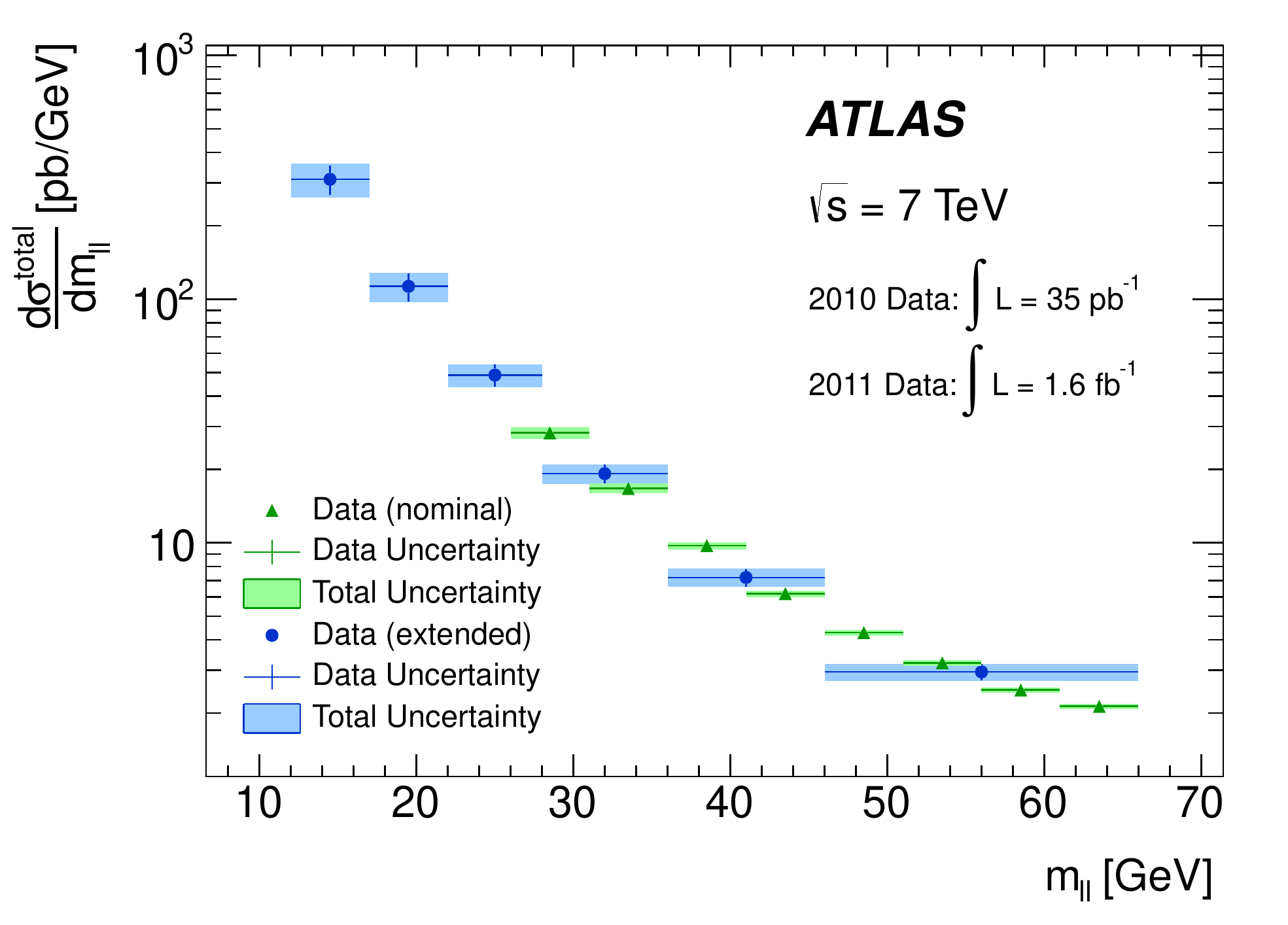}
  \caption{Comparison of Born-level nominal $\left(e+\mu\right)$
    and extended $\left(\mu\right)$ channel differential cross
    sections  as a function of the dilepton invariant
    mass, $m_{\ell\ell}$, extrapolated to full phase space. The data uncertainties
    are the total fiducial cross-section uncertainties, while the
    total uncertainties also include theoretical uncertainties from
    the acceptance correction. The luminosity uncertainties (nominal
    1.8\%, extended 3.5\%) are included in the error band.
    \label{figure:CombinedExtrapolatedCrossSection}}
\end{figure}

The measurements of the nominal and extended analyses cannot be
compared directly due to the different fiducial regions. A comparison
of the Born-level extrapolated measurements, ${\rm
  d}\sigma^\text{total}/{\rm d}m_{\ell\ell}$, determined by application of the acceptance
correction factors is shown in
figure~\ref{figure:CombinedExtrapolatedCrossSection}.
The  two measurements are in good agreement with each other and show
the expected rapid decrease of the cross section with increasing $m_{\ell\ell}$.

\subsection{Theory comparison}

The fiducial cross-section measurements are compared to theoretical
predictions from {\sc Fewz} at NLO and NNLO as well as NLO
calculations matched to a LL resummed parton shower calculation from {\sc
  Powheg}. In order to compare the QCD calculations to the
data, additional corrections are required to account for higher-order
electroweak radiative effects~\cite{Dittmaier:2009cr} and photon
induced processes, $\gamma\gamma\rightarrow \ell\ell$
~\cite{photonInducedCorrections}. The calculations are performed using
{\sc Fewz} and cross checked with {\sc Sanc}~\cite{SANC1}.

The electroweak corrections calculated in the $G_{\mu}$ scheme, $\Delta^\text{HOEW}$, account for the effects
of pure weak-vertex and self-energy corrections, double boson
exchange, initial-state radiation (ISR), and the interference between
ISR and FSR. A comparison of the HOEW corrections obtained with the
alternative $\alpha(M_Z)$ electroweak scheme~\cite{Beringer:1900zz}, $\Delta^\text{HOEW}_{\alpha(M_Z)}$,
yields different results at low $m_{\ell\ell}$ and the difference,
$\delta^{scheme}$, is listed
in tables~\ref{table:nominalCorrections}
and~\ref{table:extendedCorrections}, where
$\Delta^{\rm HOEW} = \Delta^{\rm HOEW}_{\alpha(M_Z)} - \delta^{scheme}$.  


The cross-section contribution from photon induced processes,
$\Delta^\text{PI}$, is estimated using the MRST2004QED PDF
set~\cite{MRST2004qed} in which photon radiation from the quark lines
is included in the parton evolution equations. The cross-section predictions are
calculated using the NLO and NNLO {\sc MSTW2008} sets as appropriate.
The full cross-section predictions including all corrections are shown in tables
~\ref{table:nominalTheory} and ~\ref{table:extendedTheory} for nominal
and extended analysis respectively. The corrections and associated
uncertainties are also listed in tables~\ref{table:nominalCorrections}
and ~\ref{table:extendedCorrections} for both fiducial
measurements. The $\Delta^\text{PI}$ corrections contribute $2$--$3\%$ of the
theoretical predictions.

\begin{figure}[!htb]
\singlespacing
\centering
\subfigure[\label{figure:NominalTheory}]{\includegraphics[width=0.48\textwidth]{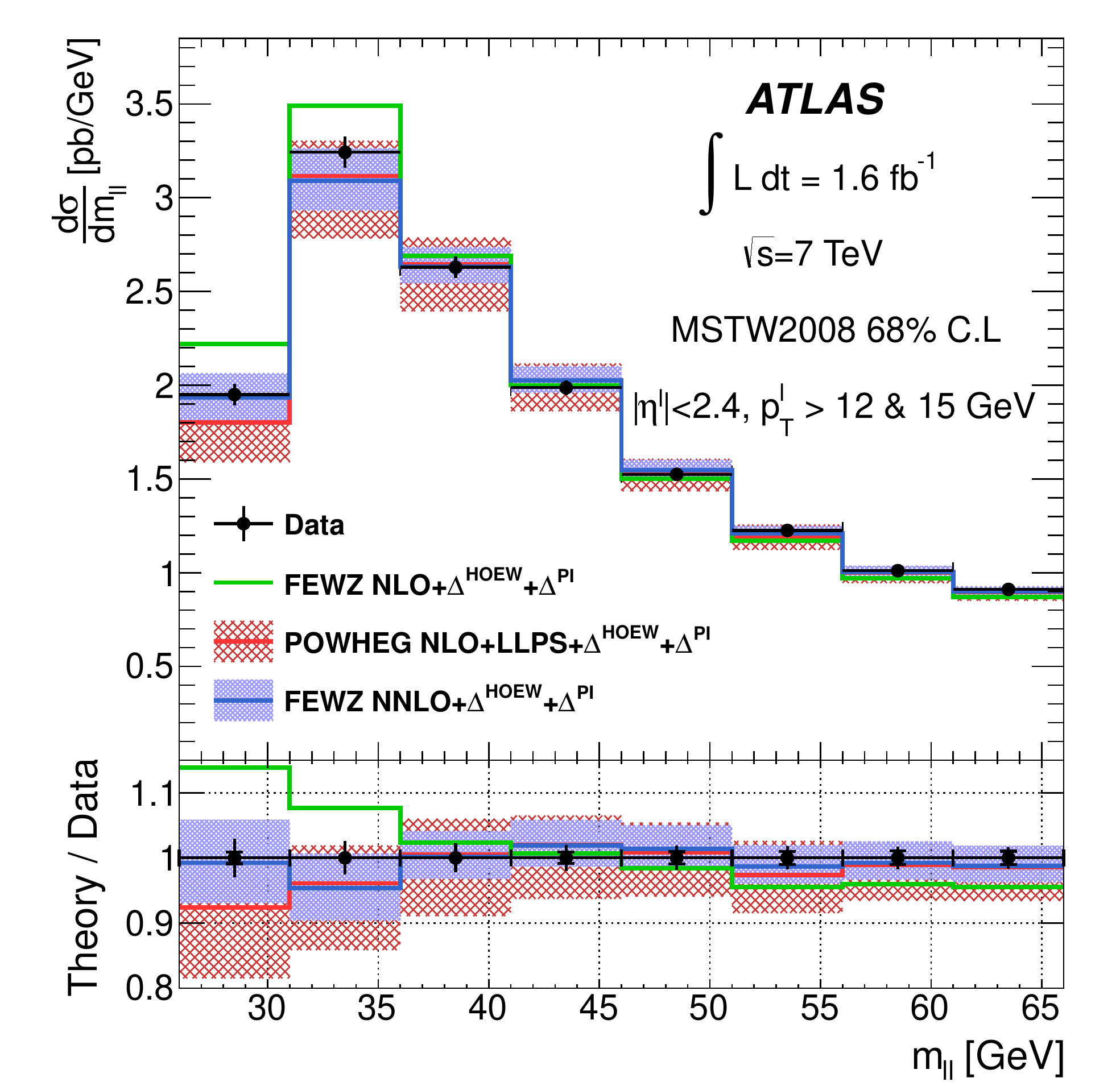}}
\subfigure[\label{figure:ExtendedTheory}]{\includegraphics[width=0.48\textwidth]{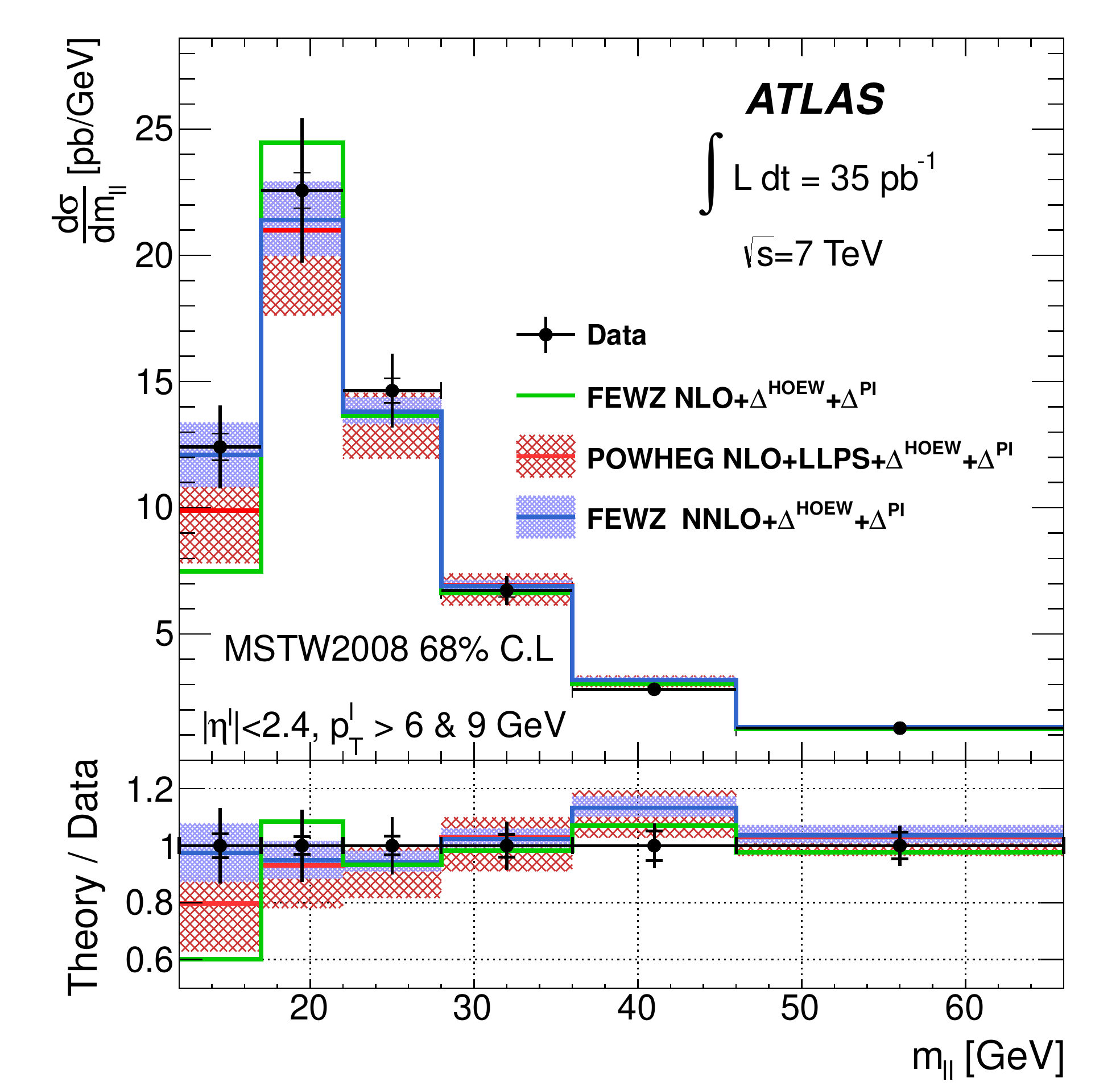}}
\caption{The measured fiducial differential cross section,
  $\frac{\text{d}\sigma}{\text{d}m_{\ell\ell}}$ for
  \subref{figure:NominalTheory} the nominal analysis and \subref{figure:ExtendedTheory}
  the extended analysis as a function of the invariant mass $m_{\ell\ell}$
  (solid points) compared to NLO predictions from {\sc Fewz}, NLO+LLPS
  predictions from {\sc Powheg} and NNLO predictions from {\sc Fewz}
  (all including higher-order electroweak and photon induced
  corrections). 
  The predictions are calculated using {\sc MSTW2008} PDF sets with the appropriate order of perturbative QCD. The uncertainty bands include the PDF and
  $\alpha_s$ variations at $68\%$ CL, scale variations between 0.5 and
  2 times the nominal scales, and the uncertainty in the PI correction.  The
  ratios of all three theoretical predictions (solid lines) to the
  data are shown in the lower panels. The data  (solid
  points) are displayed at unity with the statistical (inner) and
  total (outer) measurement
  uncertainties.}\label{figure:TheoryComparisons}
\end{figure}

The comparisons between the measured cross sections and the
theoretical predictions are shown in
figure~\ref{figure:TheoryComparisons}.  The {\sc Fewz} NLO
predictions provide a poor description of the data at low $m_{\ell\ell}$
which simultaneously overestimates and underestimates the nominal and
extended measurements respectively. The {\sc Powheg} predictions
differ from {\sc Fewz} by as much as $20\%$ and describe the data
well. These calculations have an uncertainty dominated by the scale
variations which can reach $10\%$ to $20\%$ in the lowest $m_{\ell\ell}$ bin
for each fiducial measurement. Such relatively large scale effects at
NLO can arise since the region of $m_{\ell\ell}\sim 2p^{\ell}_{\rm T}$ is only
populated by NLO type events leading to unusually large scale
variations. The {\sc Powheg} calculations absorb resummed LL parton
shower effects, which improve the prediction in this region. At NNLO
the pure fixed-order {\sc Fewz} predictions also compare well with the
measured fiducial cross sections. The associated scale uncertainties
are in this case much smaller, but still at the level of $5\%$ in the
lowest bin of nominal, and $10\%$ in the lowest bin of extended
measurements, respectively.

\begin{table}[!htb]
\begin{center}
\begin{tabular}{c|ccc|ccc|ccc}
\hline\hline
& \multicolumn{3}{|c|}{{\sc Powheg}} & \multicolumn{3}{|c}{\sc Fewz NLO}  & \multicolumn{3}{|c}{\sc Fewz NNLO}\\
$m_{\ell\ell}$  &  $\frac{\text{d}\sigma}{\text{d}m_{\ell\ell}}$  & $\delta^{\rm pdf}$ &
$\delta^{\rm scale}$ &             $\frac{\text{d}\sigma}{\text{d}m_{\ell\ell}}$  &
$\delta^{\rm pdf}$ & $\delta^{\rm scale}$ &
$\frac{\text{d}\sigma}{\text{d}m_{\ell\ell}}$ &  $\delta^{\rm pdf}$ & $\delta^{\rm scale}$   \\

  [GeV] & [pb/GeV] & [\%]&[\%]& [pb/GeV]& [\%]&[\%]  &[pb/GeV]  & [\%]  &[\%] \\
\hline       
\rule{0pt}{3ex}
26--31  &$1.80$& $2.5$ & $^{+~7.3}_{-11.4}$& $2.22$ & $2.7$  & $^{+4.9}_{ -7.9}$ & $1.93$ &$^{+3.5}_{ -2.7}$   & $5.7$  \\ \rule{0pt}{3ex}
31--36  &$3.12$& $2.4$ & $^{+~5.3}_{-10.0}$& $3.49$ & $2.7$  & $^{+4.7}_{ -6.3}$ & $3.04$ &$^{+3.2}_{ -2.5}$   & $4.5$  \\ \rule{0pt}{3ex}
36--41  &$2.64$& $2.3$ & $^{+4.6}_{-8.8}$ & $2.69$ & $2.6$  & $^{+4.1}_{ -5.0}$ & $2.58$ &$^{+3.1}_{ -2.4}$   & $2.3$  \\ \rule{0pt}{3ex}
41--46  &$2.03$& $2.2$ & $^{+3.5}_{-7.5}$ & $2.00$ & $2.6$  & $^{+3.6}_{ -4.2}$ & $1.98$ &$^{+3.1}_{ -2.3}$   & $2.1$  \\ \rule{0pt}{3ex}
46--51  &$1.54$& $1.9$ & $^{+3.7}_{-6.1}$ & $1.50$ & $2.5$  & $^{+3.2}_{ -3.5}$ & $1.51$ &$^{+3.0}_{ -2.2}$   & $1.7$  \\ \rule{0pt}{3ex}
51--56  &$1.19$& $2.4$ & $^{+4.5}_{-5.1}$ & $1.17$ & $2.4$  & $^{+2.8}_{ -2.9}$ & $1.18$ &$^{+2.9}_{ -2.2}$   & $1.3$  \\ \rule{0pt}{3ex}
56--61  &$1.00$& $2.4$ & $^{+2.3}_{-4.7}$ & $0.97$ & $2.4$  & $^{+2.6}_{ -2.6}$ & $0.98$ &$^{+2.9}_{ -2.1}$   & $1.3$  \\ \rule{0pt}{3ex}
61--66  &$0.90$& $2.1$ & $^{+2.0}_{-4.5}$ & $0.87$ & $2.3$  & $^{+2.3}_{ -2.3}$ & $0.88$ &$^{+2.8}_{ -2.1}$   & $1.2$  \\ [1ex]
\hline
\hline
\end{tabular}
\caption{Theory predictions for NLO+LLPS and for fixed-order
  calculations at NLO and NNLO including higher-order electroweak  corrections, for the
  nominal analysis of the differential cross section
  $\frac{\text{d}\sigma}{\text{d}m_{\ell\ell}}$ as a function of the
  invariant mass $m_{\ell\ell}$. The scale uncertainty is defined as the envelope
  of variations for $0.5\leq \mu_R,\mu_F\leq2$ for {\sc Powheg}. For
  {\sc Fewz} the scale uncertainty is defined by the variation
  $0.5\leq \mu_R=\mu_F\leq2$.}
\label{table:nominalTheory}
\end{center}
\end{table}

\begin{table}[!htb]
\begin{center}             
\begin{tabular}{c|ccc|ccc|ccc}
\hline\hline
& \multicolumn{3}{|c|}{{\sc Powheg}} & \multicolumn{3}{|c}{\sc Fewz NLO}  & \multicolumn{3}{|c}{\sc Fewz NNLO}\\
$m_{\mu\mu}$  &  $\frac{\text{d}\sigma}{\text{d}m_{\mu\mu}}$  & $\delta^{\rm pdf}$ & $\delta^{\rm scale}$ & 
                $\frac{\text{d}\sigma}{\text{d}m_{\mu\mu}}$  & $\delta^{\rm pdf}$ & $\delta^{\rm scale}$ & 
                $\frac{\text{d}\sigma}{\text{d}m_{\mu\mu}}$  & $\delta^{\rm pdf+\alpha_{s}}$  & $\delta^{\rm scale}$  \\
 
[GeV] & [pb/GeV] & [\%]&[\%]& [pb/GeV]  & [\%]&[\%]&[pb/GeV]  & [\%]  &[\%]\\
\hline                      \rule{0pt}{3ex}                                                          
 12--17 &$9.88	$ & $2.3$ & $^{+12.3}_{-20.9}$& $ 7.47$ & $2.7$ & $^{+10.7}_{-15.8}$ &  $12.09$ & $^{+3.7}_{-3.0}$  &   $10.0$    \\ \rule{0pt}{3ex}
 17--22 &$20.99$ & $2.6$ & $^{+~8.4}_{-15.6}$ & $24.46$ & $3.0$ & $^{+10.1}_{-13.3}$ &  $21.22$ & $^{+3.7}_{-2.8}$  &   $ 6.1$    \\ \rule{0pt}{3ex}
 22--28 &$13.69$ & $2.6$ & $^{+~5.5}_{-12.1}$ & $13.65$ & $2.9$ & $^{+6.2}_{-8.6}$   &  $13.56$ & $^{+3.4}_{-2.6}$  &   $ 2.3$    \\ \rule{0pt}{3ex}
 28--36 &$6.92	$ & $2.3$ & $^{+~6.2}_{-10.8}$ & $ 6.61$ & $2.7$ & $^{+5.0}_{-6.5}$   &  $ 6.74$ & $^{+3.3}_{-2.5}$  &   $ 1.3$    \\ \rule{0pt}{3ex}
 36--46 &$3.18	$ & $2.3$ & $^{+4.4}_{-8.6}$  & $ 3.01$ & $2.6$ & $^{+4.0}_{-4.4}$   &  $ 3.10$ & $^{+3.1}_{-2.3}$  &   $ 1.2$    \\ \rule{0pt}{3ex}
 46--66 &$1.31	$ & $2.2$ & $^{+2.9}_{-5.7}$  & $ 1.24$ & $2.4$ & $^{+2.8}_{-3.0}$   &  $ 1.28$ & $^{+2.9}_{-2.1}$  &   $ 1.3$    \\ [1ex]
\hline                                      
\hline                                                                                                                                 
\end{tabular}                                                                                                                          
\caption{Theory predictions for NLO+LLPS and for fixed-order
  calculations at
  NLO and NNLO including higher-order electroweak  corrections, for the extended
  analysis of the differential cross section
  $\frac{\text{d}\sigma}{\text{d}m_{\ell\ell}}$ as a function of the
  invariant mass $m_{\ell\ell}$. The scale uncertainty is defined as the envelope of
  variations for $0.5\leq \mu_R,\mu_F\leq2$ for {\sc Powheg}. For {\sc
    Fewz} the scale uncertainty is defined by the variation $0.5\leq
  \mu_R=\mu_F\leq2$.}
\label{table:extendedTheory}
\end{center}
\end{table}

\begin{table}[!htb]
\begin{center}

\begin{tabular}{c|c|c|c}
\hline\hline
\rule{0pt}{3ex}
$m_{\ell\ell}$  & $\Delta^\text{HOEW} $ &  $\Delta^\text{PI}$  & $\delta^{scheme}$  \\ 
      {[GeV]}  &           [\%]       &        [pb/GeV]     &        [\%]       \\
\hline       \rule{0pt}{3ex}
$26-31$ & $1.10$ &  $0.005 \pm 0.002$ &  $+4.6$\\  \rule{0pt}{3ex}
$31-36$ & $3.10$ &  $0.051 \pm 0.018$ &  $+1.5$\\  \rule{0pt}{3ex}
$36-41$ & $3.92$ &  $0.053 \pm 0.019$ &  $+0.8$\\  \rule{0pt}{3ex}
$41-46$ & $4.25$ &  $0.045 \pm 0.016$ &  $+0.5$\\  \rule{0pt}{3ex}
$46-51$ & $4.46$ &  $0.036 \pm 0.013$ &  $+0.4$\\  \rule{0pt}{3ex}
$51-56$ & $4.43$ &  $0.029 \pm 0.010$ &  $+0.4$\\  \rule{0pt}{3ex}
$56-61$ & $4.47$ &  $0.023 \pm 0.008$ &  $+0.3$\\  \rule{0pt}{3ex}
$61-66$ & $4.09$ &  $0.019 \pm 0.007$ &  $+0.4$\\  

\hline
\hline
\end{tabular}
\caption{Higher-order electroweak corrections in nominal analysis,
  $\Delta^\text{HOEW}$, and the correction for the Photon Induced process, $\Delta^\text{PI}$, together
  with its uncertainty derived from the uncertainty of the photon
  PDF as a function of the dilepton invariant mass $m_{\ell\ell}$.  Also shown is the difference arising from the
  non-convergence of calculations derived with different electroweak
  schema, $\delta^{scheme}$.
}
\label{table:nominalCorrections}
\end{center}
\end{table}

\begin{table}[!htb]
\begin{center}

\begin{tabular}{c|c|c|c}
\hline\hline
\rule{0pt}{3ex}
$m_{\ell\ell}$  & $\Delta^\text{HOEW} $ &  $\Delta^\text{PI}$  & $\delta^{scheme}$  \\ 
      {[GeV]}   &           [\%]        &       [pb/GeV]      &        [\%]       \\
\hline       \rule{0pt}{3ex}
$12-17$  &$0.37$ & $0.000 \pm 0.000$ & $+5.4$\\  \rule{0pt}{3ex}
$17-22$  &$1.58$ & $0.190 \pm 0.070$ & $+3.2$\\  \rule{0pt}{3ex}
$22-28$  &$3.04$ & $0.240 \pm 0.087$ & $+0.9$\\  \rule{0pt}{3ex}
$28-36$  &$3.77$ & $0.150 \pm 0.054$ & $+0.5$\\  \rule{0pt}{3ex}
$36-46$  &$4.38$ & $0.085 \pm 0.030$ & $+0.3$\\  \rule{0pt}{3ex}
$46-66$  &$4.64$ & $0.037 \pm 0.013$ & $+0.2$\\  [0.5ex]             
\hline
\hline
\end{tabular}
\caption{Higher-order electroweak corrections in extended analysis,
  $\Delta^\text{HOEW}$, and the correction for the Photon Induced process, $\Delta^\text{PI}$, together
  with its uncertainty derived from the uncertainty of the photon
  PDF as a function of the dilepton invariant mass $m_{\ell\ell}$. Also shown is the difference arising from the
  non-convergence of calculations derived with different electroweak
  schema, $\delta^{scheme}$.
}
\label{table:extendedCorrections}
\end{center}
\end{table}

To quantify the level of agreement between the measured cross sections
and the predictions, the value of the  $\chi^2$ function is calculated
taking into account the correlated experimental systematic uncertainties as well as
the theoretical uncertainties arising from the PDFs and scale variations. The
$\chi^2$ function is defined as in ref.~\cite{HERAPDF1.0} and the results are
shown in table~\ref{tab:theory}.

\begin{table}[!htb]
\begin{center}
\begin{tabular}{l|r|r}
\hline
\hline
  Prediction &  $\chi^2$ (8 points)&  $\chi^2$ (6 points) \\
             & Nominal & Extended \\
\hline
  {\sc Powheg} NLO+LLPS          & 22.4 (19.8)    & 22.3 (18.6)  \\
  {\sc Fewz} NLO                 & 48.7 (28.6)    & 139.7 (133.7)  \\
  {\sc Fewz} NNLO                & 13.9 (12.9)    & 7.1 (7.1)  \\
 \hline\hline
\end{tabular}
\caption{ Values of $\chi^2$ for the 
nominal and extended DY cross-section measurements for
predictions based from {\sc Fewz} at NLO and NNLO and {\sc Powheg}
using MSTW2008 PDFs accounting for the
experimental correlated systematic uncertainties. The values in the brackets show the $\chi^2$ when the 
PDF and scale uncertainties on the theoretical predictions are taken into account. }
\label{tab:theory}
\end{center}
\end{table}

\begin{table}[!htb]
\begin{center}
\begin{tabular}{l|r|r}
\hline
\hline
  Prediction &  $\chi^2$ (8 points)&  $\chi^2$ (6 points) \\
             & Nominal & Extended \\
\hline
NLO Fit              & 40.7    & 117.1  \\
NNLO Fit             & ~8.5    & ~~7.8  \\
\hline\hline
\end{tabular}
\caption{The $\chi^2$ values from NLO and NNLO QCD fits
  made using the ATLAS nominal and extended Drell--Yan cross-section
  measurements and the HERA-I dataset accounting for the experimental
  correlated systematic uncertainties.}
\label{tab:qcdfit} 
\end{center}
\end{table}
The values of the $\chi^2$ function obtained with
MSTW2008 PDFs are good when compared to {\sc Powheg} or {\sc
  Fewz} at NNLO; however the {\sc Fewz} NLO prediction yields very large
values. Thus, the measured cross sections are significantly more compatible with the NNLO
prediction than with the NLO prediction.

To investigate to what extent the disagreement with a pure NLO
calculation is dependent on the PDF used, a QCD analysis of the data is
performed. In this analysis the $\chi^2$ function is evaluated after
fitting the PDFs to deep inelastic scattering data from
HERA~\cite{HERAPDF1.0} and the new measurements presented here. The
QCD fit is performed using MCFM~\cite{mcfm} and
{\sc ApplGrid}~\cite{applgrid} in the HERAFitter
framework~\cite{HERAPDF1.0,heraFitter2,herafitter3,herafitter4,QCDNUM,minuit,thorne1,thorne2} at NLO and at NNLO using
additional NNLO K-factors obtained from {\sc Fewz}. The PDFs are
parameterised using functional forms described in
ref.~\cite{atlas-strange} where terms are added in the polynomial
expansion of the PDFs only if required by the data, following the
procedure described in ref.~\cite{HERAPDF1.0}. The data are
then included in the $\chi^2$ function, which is minimised with respect
to the PDF parameters. The NLO and NNLO QCD fits using only HERA data
yield acceptable fits with $\chi^2/\text{dof}=468.3/537$ and $\chi^2/\text{dof}=466.3/537$ respectively. When the new ATLAS measurements are included the NLO
fit is unable to describe the data and the total $\chi^2$ value for
these measurements, taking into account $20$ sources of correlated
uncertainty, is $158$ for $14$ data points. However, the QCD analysis performed at NNLO
results in a good fit with a total $\chi^2$ value of $16.3$ for $14$
measurements.

\begin{figure}[!htb]
\begin{center}
\subfigure[\label{fig:qcdfit_data_nominal}]{\includegraphics[width=0.48\linewidth]{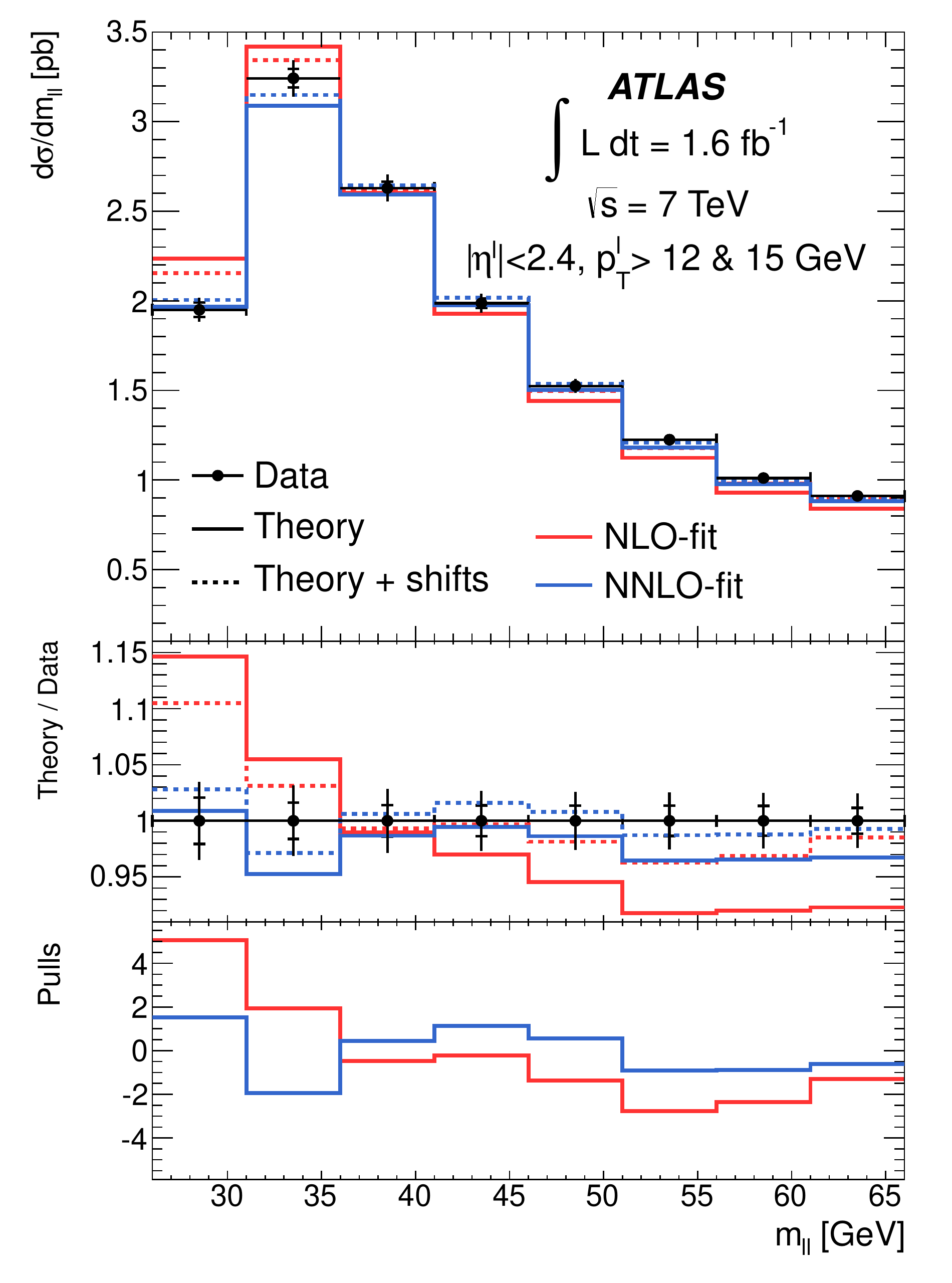}}
\subfigure[\label{fig:qcdfit_data_extended}]{\includegraphics[width=0.48\linewidth]{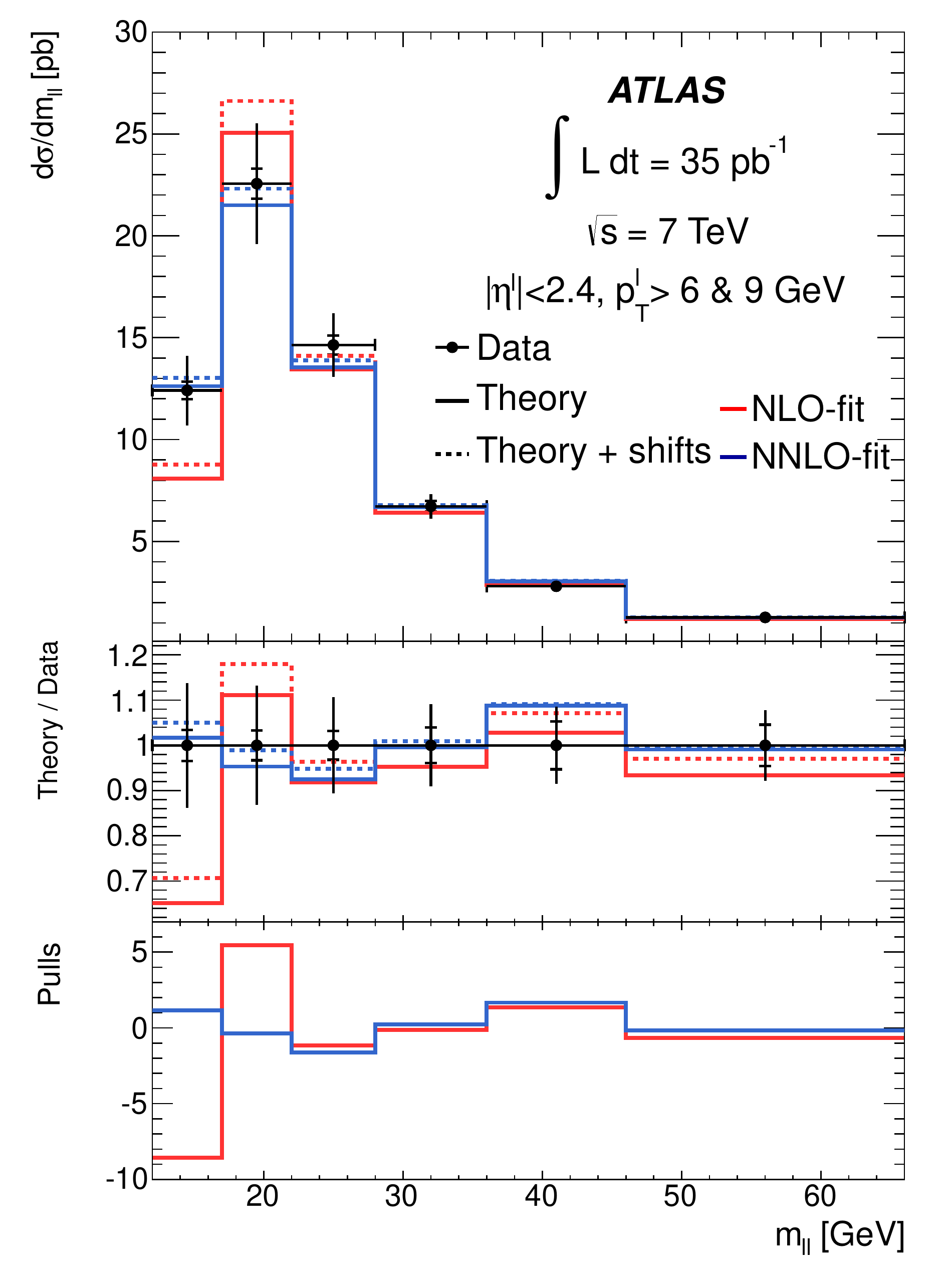}}
\caption{\label{fig:qcdfit_data}
The measured differential cross section $\frac{\text{d}\sigma}{\text{d}m_{\ell\ell}}$ for ~\subref{fig:qcdfit_data_nominal} the nominal
and~\subref{fig:qcdfit_data_extended}  the extendedanalysis as a
function of invariant mass $m_{\ell\ell}$ compared to the NLO and NNLO QCD fits (solid lines). The inner error bars show the total
uncorrelated experimental uncertainty, and the outer error bars represent
the total experimental uncertainty, excluding the luminosity
uncertainties. The dashed lines correspond to the QCD fit after
applying the adjustments of the fitted nuisance parameters for
each correlated error source. The lower half of each figure shows the
ratio of
theory expectations to data  in the upper part, and the $\chi^2$ pull contribution in the lower part.}
\end{center}
\end{figure}

The results of the
analysis are given in table~\ref{tab:qcdfit} and
figure~\ref{fig:qcdfit_data}
where the NLO and NNLO QCD fit results are shown in the upper panels and
compared to the nominal and extended analysis measurements.
Also shown are the results of the fit after applying the adjustments 
of the fitted nuisance parameters for each correlated error source. The central
panels shows the ratio of theory to data, and the lower panel shows
the $\chi^2$ pull for each measurement bin. The pulls for the NLO fit
are large at low invariant masses as expected from the phase space
constraints created by the transverse momentum cuts applied to the
leptons. However,  there are also large pulls for the higher
invariant masses.
%

\section{Conclusion}
\label{section:Conclusion}
The differential cross section ${\rm d}\sigma/{\rm  d}m_{\ell\ell}$ for the Drell--Yan production of dileptons is
measured by ATLAS in $pp$ collisions at $\sqrt{s}=7$~TeV at the LHC. The
measurements are performed using di-electron and di-muon events in a nominal fiducial region using
$1.6$~fb$^{-1}$ of integrated luminosity from 2011 covering the range
$26<m_{\ell\ell}<66$~GeV and $|\eta|<2.4$. The measurement is also
performed using di-muon events in an extended
fiducial region covering the range $12<m_{\ell\ell}<66$~GeV  and
$|\eta|<2.4$ using
$35$~pb$^{-1}$ of integrated luminosity from 2010. 

The fiducial cross sections are compared to fixed order theoretical
predictions at NLO and NNLO from {\sc Fewz}, as well as an NLO with
matched leading-logarithm parton shower calculation from {\sc
  Powheg}. The calculations are corrected for additional higher-order
electroweak radiative effects including a photon induced term.  Using
renormalisation and factorisation scales equal to $m_{\ell\ell}$, the
pure NLO prediction using the MSTW2008 PDF yields a very large
$\chi^2$ value, whereas both the NNLO and NLO matched to leading
logarithm parton shower predictions provide good descriptions of the
data. The results are supported by a QCD analysis of the measurements
performed at NLO and NNLO. The PDFs are fitted to the new measurements
together with inclusive $ep$ measurements from HERA. The NNLO fit
performs significantly better than the NLO fit in describing the data.

\clearpage
\section{Acknowledgements}


We thank CERN for the very successful operation of the LHC, as well as the
support staff from our institutions without whom ATLAS could not be
operated efficiently.

We acknowledge the support of ANPCyT, Argentina; YerPhI, Armenia; ARC,
Australia; BMWF and FWF, Austria; ANAS, Azerbaijan; SSTC, Belarus; CNPq and FAPESP,
Brazil; NSERC, NRC and CFI, Canada; CERN; CONICYT, Chile; CAS, MOST and NSFC,
China; COLCIENCIAS, Colombia; MSMT CR, MPO CR and VSC CR, Czech Republic;
DNRF, DNSRC and Lundbeck Foundation, Denmark; EPLANET, ERC and NSRF, European Union;
IN2P3-CNRS, CEA-DSM/IRFU, France; GNSF, Georgia; BMBF, DFG, HGF, MPG and AvH
Foundation, Germany; GSRT and NSRF, Greece; ISF, MINERVA, GIF, I-CORE and Benoziyo Center,
Israel; INFN, Italy; MEXT and JSPS, Japan; CNRST, Morocco; FOM and NWO,
Netherlands; BRF and RCN, Norway; MNiSW and NCN, Poland; GRICES and FCT, Portugal; MNE/IFA, Romania; MES of Russia and ROSATOM, Russian Federation; JINR; MSTD,
Serbia; MSSR, Slovakia; ARRS and MIZ\v{S}, Slovenia; DST/NRF, South Africa;
MINECO, Spain; SRC and Wallenberg Foundation, Sweden; SER, SNSF and Cantons of
Bern and Geneva, Switzerland; NSC, Taiwan; TAEK, Turkey; STFC, the Royal
Society and Leverhulme Trust, United Kingdom; DOE and NSF, United States of
America.

The crucial computing support from all WLCG partners is acknowledged
gratefully, in particular from CERN and the ATLAS Tier-1 facilities at
TRIUMF (Canada), NDGF (Denmark, Norway, Sweden), CC-IN2P3 (France),
KIT/GridKA (Germany), INFN-CNAF (Italy), NL-T1 (Netherlands), PIC (Spain),
ASGC (Taiwan), RAL (UK) and BNL (USA) and in the Tier-2 facilities
worldwide.



\newpage
\bibliographystyle{JHEP}
\bibliography{DYpaper}{}

\onecolumn
\clearpage
\begin{flushleft}
{\Large The ATLAS Collaboration}

\bigskip

G.~Aad$^{\rm 84}$,
T.~Abajyan$^{\rm 21}$,
B.~Abbott$^{\rm 112}$,
J.~Abdallah$^{\rm 152}$,
S.~Abdel~Khalek$^{\rm 116}$,
O.~Abdinov$^{\rm 11}$,
R.~Aben$^{\rm 106}$,
B.~Abi$^{\rm 113}$,
M.~Abolins$^{\rm 89}$,
O.S.~AbouZeid$^{\rm 159}$
% ...
\bigskip
\\
% ...
$^{11}$ Institute of Physics, Azerbaijan Academy of Sciences, Baku, Azerbaijan\\
% ...
$^{21}$ Physikalisches Institut, University of Bonn, Bonn, Germany\\
% ...
$^{84}$ CPPM, Aix-Marseille Universit{\'e} and CNRS/IN2P3, Marseille, France\\
% ...
$^{89}$ Department of Physics and Astronomy, Michigan State University, East Lansing MI, United States of America\\
% ...
$^{106}$ Nikhef National Institute for Subatomic Physics and University of Amsterdam, Amsterdam, Netherlands\\
% ...
$^{112}$ Homer L. Dodge Department of Physics and Astronomy, University of Oklahoma, Norman OK, United States of America\\
$^{113}$ Department of Physics, Oklahoma State University, Stillwater OK, United States of America\\
% ...
$^{116}$ LAL, Universit{\'e} Paris-Sud and CNRS/IN2P3, Orsay, France\\
% ...
$^{152}$ Institute of Physics, Academia Sinica, Taipei, Taiwan\\
% ...
$^{159}$ Department of Physics, University of Toronto, Toronto ON, Canada\\
% ...
\end{flushleft}


\end{document}